\documentclass[fleqn,usenatbib]{mnras}

\usepackage{newtxtext,newtxmath}

\usepackage[T1]{fontenc}

\DeclareRobustCommand{\VAN}[3]{#2}
\let\VANthebibliography\thebibliography
\def\thebibliography{\DeclareRobustCommand{\VAN}[3]{##3}\VANthebibliography}

\usepackage{graphicx}	
\usepackage{amsmath}	
\usepackage{orcidlink}
\usepackage{multirow}

\newcommand{\ark}{\textsc{Arkenstone}}
\defcitealias{Smith2024a}{S24}

\title[Prevention is better than cure?]{Prevention is better than cure? Feedback from high specific energy winds in cosmological simulations with \textsc{Arkenstone}}

\author[Bennett et al.]{%
Jake S. Bennett\orcidlink{0000-0002-8573-2993},$^{1}$\thanks{E-mail: jake.bennett@cfa.harvard.edu}
Matthew C. Smith\orcidlink{0000-0002-9849-877X},$^{2}$
Drummond B. Fielding\orcidlink{0000-0003-3806-8548},$^{3,5}$
Greg L. Bryan\orcidlink{0000-0003-2630-9228},$^{4,5}$
\newauthor
Chang-Goo Kim\orcidlink{0000-0003-2896-3725},$^{6}$
Volker Springel\orcidlink{0000-0001-5976-4599},$^2$ 
Lars Hernquist,$^1$
Rachel S. Somerville\orcidlink{0000-0002-6748-6821},$^5$
and Laura Sommovigo\orcidlink{0000-0002-2906-2200}$^5$
\vspace*{0.1cm}\\%
$^{1}$Center for Astrophysics | Harvard \& Smithsonian, 60 Garden Street, Cambridge, MA 02138, USA\\%
$^{2}$Max-Planck-Institut f{\"u}r Astrophysik, Karl-Schwarzschild-Str. 1, D-85748, Garching, Germany\\%
$^{3}$Department of Astronomy, Cornell University, Ithaca, NY 14853, USA\\%
$^{4}$Department of Astronomy, Columbia University, 550 West 120\textsuperscript{th} Street, New York, NY 10027, USA\\%
$^{5}$Center for Computational Astrophysics, Flatiron Institute, 162 5\textsuperscript{th} Avenue, New York, NY 10010, USA\\%
$^{6}$Department of Astrophysical Sciences, Princeton University, 4 Ivy Lane, Princeton, NJ 08544, USA\\%
}

\date{MNRAS, published}

\pubyear{2025}

\begin{document}
\label{firstpage}
\pagerange{\pageref{firstpage}--\pageref{lastpage}}
\maketitle

\begin{abstract}
We deploy the new \textsc{Arkenstone} galactic wind model in cosmological simulations for the first time, allowing us to robustly resolve the evolution and impact of high specific energy winds. 
In a (25\,$h^{-1}$\,Mpc)$^3$ box we perform a set of numerical experiments that systematically vary the mass and energy loadings of such winds, finding that their energy content is the key parameter controlling the stellar to dark matter mass ratio. Increasing the mass loading, at fixed energy, actually results in mildly {\it enhanced} star formation, counter to prevailing wisdom, due to the wind becoming cooler.  
Of the simple parametrisations that we test, we find that an energy loading that scales inversely with halo mass best matches a wide range of observations and can do so with mass loadings drastically lower than those in most previous cosmological simulations.
In this scenario, much less material is ejected from the interstellar medium. Instead, winds both heat gas in the circumgalactic medium, slowing infall onto the galaxy, and also drive shocks beyond the virial radius, decreasing the halo-scale accretion rate.
We can also report that a much lower fraction of the available supernova energy is needed in preventative galaxy regulation than required by ejective wind feedback models such as IllustrisTNG. This is a Learning the Universe collaboration publication.
\end{abstract}




\section{Introduction} \label{Section:Intro}

In the $\Lambda$CDM cosmological model, galaxy formation is driven by the gravitational collapse of dark matter (DM) hierarchically into DM haloes, with subsequent condensation of baryons into their gravitational potential followed by the formation of stars \citep{ReesOstriker1977,Silk1977,WhiteRees1978,WhiteFrenk1991}. Our modern picture of galaxy formation involves a complex interplay of accretion onto galaxies and their interstellar medium (ISM), along with outflows of mass and energy driven by feedback processes. This feedback is thought to be essential, as simulations that do not include it fail to reproduce many fundamental observations of galaxy properties, converting too large a fraction of baryons into stars and producing galaxies that are overly compact and bulge-dominated \citep[e.g.][]{Navarro1994,Navarro1995}.  

Feedback is generally believed to come from two primary sources: black holes (BHs) and stars \citep[see e.g.][]{HeckmanBest2023}. In lower mass haloes, below approximately $10^{12}$\,M$_\odot$, the dominant effect is thought to be stellar feedback, from stellar winds, radiation, and supernovae, while massive galaxies are regulated by BH feedback \citep[see e.g.][]{SomervilleDave2015}. At all mass scales, regulation of star formation occurs ultimately by reducing the efficiency of the conversion of gas into stars. On small spatial scales, energy injection causes heating and drives turbulence in the ISM, disrupting star formation \citep[e.g.,][]{Ostriker2010,Hopkins2011,FaucherGiguere2013,Krumholz2018,OstrikerKim2022}. The key to the long-term regulation of star formation however lies in the larger scale baryon cycle, in which galactic winds play a key role \citep[e.g.,][]{Strickland2000,Oppenheimer2008,Hopkins2012,AnglesAlcazar2017b,Wright2024}. 

Winds can regulate galaxy formation in two ways. They can remove material from the ISM, reducing the amount of gas available to form stars and carrying mass and metals out into the circumgalactic medium (CGM) -- \textit{ejective} feedback. They can also slow or stop gas from accreting onto the galaxy to replenish the fuel for star formation, via the injection of energy into the CGM (or beyond) -- \textit{preventative} feedback. We can characterise winds with two key parameters: the mass loading ($\eta_{\rm M}$, the ratio of the wind mass flux and the star formation rate) and the energy loading ($\eta_{\rm E}$, the ratio of the wind energy flux and the rate at which energy is released during star formation and stellar evolution). We will discuss these parameters extensively in this work.

A wealth of observational evidence exists for outflowing material around many galaxies \citep[see][for a review]{Rupke2018}, including from spectroscopic detections of outflows themselves and metal pollution of the CGM at large distances from the central galaxy \citep[e.g.][]{Prochaska2017,Guo2020,Zabl2021}. These outflows are ubiquitously found to be multiphase, with molecular, neutral and ionised gas found to co-exist using probes across the electromagnetic spectrum \citep[e.g.][]{Rupke2018,ThompsonHeckman2024}. Despite the widespread detection of outflows around galaxies, estimating the mass and energy loading values is very difficult -- especially in the hottest gas phases. Where such data exist, mass fluxes are reliably found to be comparable to or below the star formation rate \citep[e.g.][]{McQuinn2019,Marasco2022,Kado-Fong2024} and the energy flux is found to be comparable to the energy injected by supernovae \citep[e.g.][]{StricklandHeckman2009}. Though most observational estimates of mass loading are based on nearby galaxies, there have also been a number of recent studies of outflows at cosmic noon and above, which suggest a similar picture, but with differences across different gas phases and galaxy masses at higher redshift \citep{Concas2022,Cooper2025,Xu2025}.

Such multiphase winds have emerged naturally in high resolution ($\sim$pc scale) simulations that include most of the physics thought to be important to star formation and stellar feedback using small ($\sim$kpc scale) patches of the ISM \citep[e.g.][]{KimOstriker2018,Kim2020a,Kannan2020,Rathjen2023} and low mass galaxies \citep[e.g.][]{Smith2018,Smith2021b,Hu2019,Schneider2020,Schneider2024,Steinwandel2024a}. A recent consensus of such detailed simulations is that not all wind phases are created equally -- a hot, fast wind phase ($\gtrsim\!10^6$\,K) carries most of the energy in the wind, and a slower, cooler phase ($\lesssim\!10^4$\,K) carries most of the mass \citep{Fielding2018,Kim2020a,LiBryan2020}. 

Recent analytic and semi-analytic models incorporating energy injection and heating of the CGM have highlighted the importance of the hot wind phase, in the form of high specific energy winds \citep{Carr2023,Pandya2023,Voit2024a,Voit2024b}. In particular, the regulator model of \citet{Carr2023} suggested that the amount of energy carried by a hot wind is more important in regulating star formation than the amount of mass comprising the wind. However, due to the large spatial extent and timescales involved, following the evolution of such a high specific energy wind after it has launched, and the long-term effect on its host galaxy ($\sim$Mpc scales), has yet to be done in full cosmological simulations. 

To study the impact of energy and mass loadings for a statistically robust sample of galaxies, we require cosmological simulations of a fairly large volume ($\gg$Mpc scales). However, the resolution of such simulations precludes the ab initio modelling of galactic winds. Primarily this is due to under-resolving the ISM, where energy injected into dense gas is typically radiated away unphysically rapidly before an outflow can be launched. Subgrid models used for stellar feedback generally circumvent this in one of two ways. The first is by overpowering cooling through the rare, stochastic injection of large amounts of energy representing multiple supernovae events \citep[e.g.][]{DallaVecchiaSchaye2012,EAGLE1,McCarthy2017}. However, low resolution in the ISM combined with discrete injections of energy (e.g.~from supernovae) lead to a wind energy injection rate that is dependent on mass resolution -- trying to power a high specific energy wind with coarse resolution wind particles results in an artificially bursty energy injection. The second is through bypassing the ISM entirely, via launching hydrodynamically decoupled wind particles that typically recouple in the inner CGM \citep[e.g.][]{SpringelHernquist2003,Illustris1,Dave2016,Dave2019,FABLE1,Pillepich2018}. In most simulations, an effective equation of state model \citep[e.g.][]{SpringelHernquist2003,SchayeDallaVecchia2008} is used to capture the unresolved small-scale heating and turbulence within the ISM. Many existing simulations also tend to use a higher amount of feedback energy that is typically expected to come from supernovae to be able to regulate star formation, particularly in low mass galaxies (see e.g. Section~\ref{Section:Loadings}). 

Most subgrid models launch a single wind phase, with the same launch velocity and mass loading \citep[e.g.][]{Illustris1,Pillepich2018}, though several also add a thermal component to either all wind particles \citep{Pillepich2018} or a fraction of particles \citep{Dave2016,Dave2019} to mimic the impact of hot wind material. No cosmological simulation reliably resolves the interaction between hot and cold phases of material in a wind, which would likely require sub-solar mass resolution \citep[e.g.,][]{McCourt2018,GronkeOh2020a,Abruzzo2024}. An additional class of simulations attempts to resolve more details of the multiphase nature of the ISM in cosmological zoom simulations and to self-consistently track the interaction of stellar feedback and the ISM in launching outflows \citep[e.g.][]{Hopkins2014,Hopkins2018,Muratov2015,Marinacci2019,Pandya2021}. Such simulations are difficult to scale up to large cosmological volumes, though some have been used to calibrate loadings in larger-scale simulations \citep[e.g.][]{Dave2016}.

Most modern cosmological box simulations tune the parameters of their models to match particular observables (e.g. the stellar mass -- halo mass relation). After this process, the mass and energy loading factors for stellar feedback in most cosmological simulations are typically much higher than what is measured in resolved ISM simulations \citep[e.g.][]{LiBryan2020}. Different simulations can have drastically different predictions for secondary properties of galaxies, most notably in the CGM \citep{Fielding2017, JDavies2019,JDavies2020,Fielding2020b,Kelly2022,Ayromlou2023,Wright2024}. Resolution also affects the properties of the simulated CGM, further complicating the matter, though some recent progress has been made using simulations that specifically focus computational resources by adding additional resolution elements in the CGM \citep[e.g.][]{Hummels2019,Peeples2019,vandeVoort2019,Suresh2019,Bennett2020,Ramesh2024}. How many of the differences in predicted galaxy and CGM properties between simulations are due to the numerical implementation of subgrid models is unclear; until now it has not been possible to study the regulatory effect of high specific energy, low mass winds on a large population of galaxies.

Simulating high specific energy winds is challenging in cosmological simulations \citep[see section 2 of][for a detailed discussion]{Smith2024a}. Hot, fast winds with low mass fluxes have lower densities, leading to coarser spatial resolution in a Lagrangian code (or with a quasi-Lagrangian refinement strategy in an Eulerian scheme, as is typical in cosmological volumes). Even with the highest mass resolution of current cosmological box simulations ($\sim\!10^5$\,M$_\odot$) such a wind can become completely unresolved, with single resolution elements being larger than the galaxy launching them \citep[see fig.~1 of][]{Smith2024a}. 

\ark, a new subgrid model for multiphase winds, was developed to address these issues \citep{Smith2024a,Smith2024b}. The full model launches winds containing both a high specific energy component \citep[\textsc{Arkenstone-Hot},][henceforth \citetalias{Smith2024a}]{Smith2024a} and an unresolved cold cloud component \citep[\textsc{Arkenstone-Cold},][]{Smith2024b}, each with their own mass and energy loadings. \ark\ is designed to capture the effects of multiphase winds in large-scale cosmological simulations, with resolutions coarser than some of the high-resolution zooms that attempt to model such outflows self-consistently. We note that even for such zooms it is currently impossible to capture the interaction between wind phases, a key aspect of the full \ark\ model. With this full model, we can explore a huge area of previously unexplored parameter space of mass and energy loadings to investigate the balance between ejective and preventative feedback in galaxy formation. The ultimate aim is to use parameters calibrated with highly resolved ISM simulations \citep[e.g.][]{Kim2020b} as inputs to the \ark\ model, to remove the need to calibrate to global galaxy observations, as is commonly done currently in most cosmological simulations. This is central to the goals of the SMAUG\footnote{\url{https://www.simonsfoundation.org/flatiron/center-for-computational-astrophysics/galaxy-formation/smaug/}} (Simulating Multiscale Astrophysics to Understand Galaxies) and Learning the Universe\footnote{\url{http://learning-the-universe.org}} collaborations. 

In this first paper applying the \ark\ model to cosmological simulations we focus only on the hot component of winds, allowing us to study galaxy formation in a regime where preventative feedback dominates and little mass is ejected from the ISM. We vary the key parameters of these hot winds -- mass and energy loadings -- to study how the properties of the galaxy population respond. The paper is structured as follows: In Section~\ref{Section:Methods} we describe the \ark\ model and the simulations in this paper. Sections~\ref{Section:FirstLook} and \ref{Section: z0results} show properties of the galaxies and their gaseous haloes in our simulations at $z=0$. Section~\ref{Section: z2results} then looks at \textit{how} the feedback from \ark\ regulates star formation through preventative, rather than ejective feedback. We look at the redshift evolution of star formation in Section~\ref{Section:RedshiftEvo}, discuss comparisons to recent analytic, semi-analytic and simulation work in Section~\ref{Section:Discussion}, and summarise our results in Section~\ref{Section:Conclusions}.

\section{Methods} \label{Section:Methods}

\subsection{Simulation set-up}
The simulations in this paper use \textsc{Arepo} \citep{Arepo}, a cosmological hydrodynamical code that incorporates gravity using a TreePM scheme and solves the equations of hydrodynamics on a quasi-Lagrangian, unstructured, moving Voronoi mesh, where mesh generating points move with the fluid velocity (with small corrections in selected cells to promote a regular mesh geometry). In \textsc{Arepo}, a default refinement scheme keeps gas cells within a factor of two of a given mass resolution by merging and splitting cells, as non-zero mass fluxes between cells can lead to fluctuations in cell masses. In \ark, this is supplemented with an additional refinement scheme, as described in Section~\ref{Section:Refinement}.

We perform a number of numerical experiments using \textsc{Arkenstone-Hot} in cosmological box simulations to explore the ability of high specific energy winds to act as preventative feedback. To study the impact of these winds, we vary the input parameters of the wind -- specifically the mass and energy loadings -- and investigate changes in the population-level properties of galaxies due to these variations. We emphasise that we have \textit{not} calibrated the input parameters of the \ark\ wind model in this work to match observations, we simply test the impact of certain parameter choices motivated by analytic and semi-analytic modelling, and high resolution ISM simulations.

The cosmological simulations in this work all use the same initial conditions - a periodic box with a comoving side length of $36.9$\,Mpc. The box has a cosmology consistent with the \textit{Planck} 2015 constraints \citep{Planck2015Parameters}: $\Omega_\mathrm{m} = 0.3089$, $\Omega_\Lambda = 0.6911$, $\Omega_\mathrm{b} = 0.0486$, $h = 0.6774$, $\sigma_8 = 0.8159$, $n_\mathrm{s} = 0.9667$. Initial conditions were generated with the \textsc{N-GenIC} code \citep{Millennium,NGenIC}.

The simulations contain $512^3$ dark matter particles and are initialised with the same number of gas cells, corresponding to a dark matter particle mass of $1.24 \times 10^7 \, \mathrm{M}_\odot$ and a base resolution baryon mass of $2.32 \times 10^6 \, \mathrm{M}_\odot$. As described in Section \ref{Section:ArkenstoneWinds}, wind particles and flagged cells in \ark\ have a resolution 100 times better than the base resolution, corresponding to $2.32 \times 10^4 \, \mathrm{M}_\odot$. Gravitational softening lengths for DM and stars are 1.48\,ckpc at $z>1$, and 0.74\,pkpc at $z\leq1$. Gas softenings are adaptive, but have a minimum of 0.37\,pkpc. This is typical of large-scale cosmological boxes like TNG100 \citep{TNG1,TNG2,TNG3,TNG4,TNG5}, EAGLE \citep{EAGLE1,EAGLE2}, Horizon-AGN \citep{Dubois2014}, and \textsc{Simba} \citep{Dave2019}.

\subsection{Galaxy formation model}

Aside from the supernova--driven wind model, all simulations presented in this paper share the same physical model in every respect. This model is the same as the fiducial IllustrisTNG simulations, with the exception of magnetohydrodynamics (MHD), which is omitted in this work\footnote{The \textsc{Arkenstone} model is compatible with MHD, but for simplicity in these first simulations we use hydrodynamics only.}. We will describe the \textsc{Arkenstone} wind model in Sections~\ref{Section:ArkenstoneWinds} and \ref{Section:Refinement}, and the TNG wind model in Section~\ref{Section:TNGWinds}. First, however, we describe the physics common to all simulations \citep[see][for further details]{Illustris1,Weinberger2017a,Pillepich2018}.

We include radiative cooling in the same way as the original Illustris simulation and TNG \citep[for full details, see][]{Illustris1}. Gas cooling occurs in the presence of a spatially uniform, time-varying ionising UV background, with corrections for self-shielding in dense gas \citep{FG_UVB,Rahmati2013}. The background radiation field is also locally modified by a contribution from nearby AGN \citep[see][]{Illustris1}. Cooling is included for both primordial species \citep{Cen1992,Katz1996}, and metals \citep[included using tabulated values as a function of metallicity, density, temperature, and redshift,][]{SmithB2008,Wiersma2009a}. 

The resolution achievable in the ISM within cosmological simulations precludes the explicit modelling of a multiphase medium. We therefore adopt a modified version of the effective equation of state (eEoS) model from \citet{SpringelHernquist2003} to capture the unresolved, local effects of feedback on the ISM. The modifications are the same as in Illustris and TNG \citep[see][]{Illustris1}; to avoid overpressurising the ISM, we interpolate between the \citet{SpringelHernquist2003} eEoS and an isothermal EoS at 10$^4$\,K, with a thirty per cent contribution from the eEoS. All gas cells with a number density above a threshold $n_{\rm H, 0} \approx 0.1\,\mathrm{cm}^{-3}$ join the eEoS. We note that while the pressurisation of the ISM uses a portion of the energy nominally available from supernovae, the uncertainties in both the \citet{SpringelHernquist2003} model and total amount of energy that supernovae can actually provide mean that, as in TNG, we treat this separately from the energy used to launch winds. As we discuss in Section \ref{Section:Loadings}, \ark\ uses considerably less energy than TNG to launch winds, and uses a total energy well within plausible limits for the energy released from supernovae. Metal enrichment occurs via the release of metals to neighbouring cells of star particles in the ISM, at a rate determined by stellar population ages \citep[see][and the references therein]{Illustris1,Pillepich2018}.

We use the same model for BH seeding, growth, and feedback as in TNG without modification \citep[for full details, see][]{Weinberger2017a}. BHs of mass $1.18\times10^6\,$M$_\odot$ are seeded into haloes with a mass larger than $7.38\times10^{10}\,$M$_\odot$. They can grow via mergers, or through accreting gas at the Eddington-capped Bondi accretion rate \citep{HoyleLyttleton1939,BondiHoyle1944,Bondi1952}. Feedback energy is injected as a thermal dump local to the BH at high accretion rates, and as kinetic `kicks' in the low accretion rate regime, motivated by expectations for quasar-driven winds and advection-dominated inflow-outflow solution (ADIOS) winds, respectively. We note that the interaction between \ark\ and the TNG BH accretion and feedback models will not be trivial, as the parameters of the BH model were calibrated alongside the original TNG wind model. The different stellar feedback prescriptions may therefore affect BH growth. For this reason we restrict the bulk of our analysis in this paper to halo masses for which black holes do not strongly affect the evolution of galaxies (below halo masses of $\sim\!10^{12.5}\,$M$_\odot$), and we defer a detailed study of the combined feedback effect to future work. 

As in TNG, we use a friends-of-friends algorithm \citep{FOF} and the \textsc{Subfind} halo finder \citep{Subfind1,Subfind2} to identify haloes and galaxies on-the-fly.

\subsection{TNG winds} \label{Section:TNGWinds}

We now summarise the key parts of the TNG wind model \citep[fully described in][]{Pillepich2018}. This model uses hydrodynamically decoupled wind particles, launched isotropically from star forming gas cells. TNG uses a density threshold for recoupling (5\% of the star formation threshold), at which point the wind particle dumps all of its mass, energy, momentum and metals into its host cell (referred to as simple recoupling in the \ark\ scheme).

Unlike in \ark, the wind particles in TNG radiatively cool while hydro-decoupled, with their effective density adopted from the nearest gas cell. We omit this in \ark\ for three reasons. Firstly, for high specific energy winds (which \textsc{Arkenstone-Hot} is designed to treat), the density of the local ISM, as measurable in the simulation, is not relevant -- we are implicitly assuming that the hot, fast wind is propagating through unresolved low density channels and superbubble breakouts. Secondly, for these hot winds, the radiative cooling time is too long for this additional cooling to be important. Finally, our philosophy with \ark\ is to avoid modelling the evolution of the wind within the ISM as much as possible, instead focusing on the properties of the wind outside of the ISM; radiative cooling while in the ISM could be folded into the input energy loadings if desired.

TNG uses a slightly different parametrisation of input wind properties to our default approach with \ark. Instead of choosing the energy and mass content of the wind, with specific energy as a dependent variable, TNG chooses the energy content and velocity of the wind, with mass loading being a dependent variable and thermal energy being a fixed fraction of the kinetic energy. The launch velocity kick scales with the local one-dimensional (1D) DM velocity dispersion, with two modifications -- a redshift dependence and a velocity floor -- such that the fiducial velocity kick is
\begin{equation}
    \Delta v_{\rm w, TNG} = \mathrm{max} \left[7.4\times \sigma_\mathrm{DM,1D} \left( \frac{H_0}{H(z)} \right)^{1/3}, 350\,\mathrm{km\,s}^{-1} \right].
\end{equation}
The mass loading is then set as,
\begin{equation} \label{Eqn: EtaM_TNG}
    \eta_{\rm M} = \frac{2}{(\Delta v_{\rm w, TNG})^2} \left(1 - \tau_{\rm w} \right) \eta_{\rm E} u_\star        ,
\end{equation}
where $\tau_{\rm w} = 0.1$ is the thermal fraction of the wind energy. Note that here we have expressed Equation~\ref{Eqn: EtaM_TNG} with a similar parametrisation to our Equations~\ref{Eqn: LaunchV} and \ref{Eqn: LaunchU} in the subsequent section. The combination $u_\star \eta_{\rm E}$ is equivalent to the quantity $e_w$ in equation~3 of \citet{Pillepich2018}, which has a metallicity dependence (higher wind energy at lower metallicity). \cite{Pillepich2018} also assumes a SN energy per unit stellar mass formed that is 13\% higher than our reference $u_\star$ (relating to minor differences in IMF cutoff masses). Since $u_\star$ and $\eta_{\rm E}$ are degenerate as far as the actual wind energy input is concerned (and the difference is slight), for the purpose of comparing energy loadings between \ark\ and TNG on an even footing, we fix $u_\star$ here to our value and absorb the difference into the energy loading (along with the metallicity
dependence). 

The TNG model initially forms wind particles with the same metallicity as their birth cell, but then removes metal mass from wind particles and redistributes this among neighbouring ISM cells before being kicked, preventing the removal of too many metals due to TNG's high mass loading (see Section \ref{Section:Loadings} and Appendix \ref{Section:MetalAppendix}).

Finally, we emphasise that the parameters of the TNG model were calibrated to reproduce a set of observables, namely the SFR density as a function of $z$, the galaxy stellar mass function (SMF), the stellar mass to halo mass relation (SMHM), the BH mass to stellar mass relation, halo gas fractions, and galaxy stellar sizes \citep{Pillepich2018}. As of yet, we have not undertaken a similar calibration with \ark. In this paper we show a set of numerical experiments showing the impact of changes in mass and energy loadings. A full flagship cosmological simulation, with parameters either calibrated or informed by idealised high-resolution ISM simulations, will be presented in a future work.

\subsection{\ark} \label{Section:ArkenstoneWinds}

\ark\ is a framework for launching multiphase winds in cosmological simulations. It also uses hydrodynamically decoupled wind particles to source a wind just outside the ISM, as the complex interaction between winds and the ISM is largely unresolved in coarse resolution cosmological simulations (as described in Section \ref{Section:Intro}). The use of hydrodynamically decoupled wind particles has the added benefit of allowing us to control the properties of galactic winds more closely, so we can more easily investigate the impact of these properties on the CGM. We note that \ark\ uses a completely independent wind particle implementation to that used in TNG, and is fully compatible with all other TNG physics.

In this work we only employ the \textsc{Arkenstone-Hot} framework, which launches high specific energy winds (described in \citetalias{Smith2024a}), and do not include the impact of cold clouds as further described in \citet{Smith2024b}. We briefly summarise the \textsc{Arkenstone-Hot} model below, for full details see \citetalias{Smith2024a}.

ISM material is converted into wind material at a rate of
\begin{equation}
    \dot{m}_\mathrm{w} = \eta_\mathrm{M} \dot{m}_\star,
\end{equation}
where $\dot{m}_\star$ is the star formation rate (SFR) of the cell, and $\eta_\mathrm{M}$ is the mass loading factor, which is a free parameter that we test in this work. This rate is then stochastically sampled to determine how many wind particles to launch. 

Upon creation, each wind particle is given a ``launch'' energy in addition to the conserved quantities it receives from its parent cell. In contrast to \citetalias{Smith2024a}, we launch wind particles isotropically, like in TNG, with each new particle given a kick in a random direction. We note that these velocity kicks are actually added onto the velocity of the birth cell of the wind particle in both TNG and \ark. The launch energies correspond to a velocity kick with magnitude
\begin{equation} \label{Eqn: LaunchV}
    \Delta v_\mathrm{w} = \sqrt{\frac{2 \eta_\mathrm{E, kin}}{\eta_\mathrm{M}} u_\star},
\end{equation}
and a specific internal energy of
\begin{equation} \label{Eqn: LaunchU}
    u_\mathrm{w} = \frac{\eta_\mathrm{E, th}}{\eta_\mathrm{M}} u_\star,
\end{equation}
where $\eta_\mathrm{E, kin}$ and $\eta_\mathrm{E, th}$ are the kinetic and thermal energy loadings, respectively. These are treated as free parameters that we vary in this study. Throughout this paper, we assume that the Mach number of the wind at launch is fixed at a value $\mathcal{M} = 1$, which fixes the ratio $\eta_\mathrm{E, kin}/\eta_\mathrm{E, th} = 5/9$. Like \citetalias{Smith2024a}, we adopt $u_\star = 5.26 \times 10^5 \, \mathrm{(km \, s^{-1})^2}$ as the characteristic specific energy associated with stellar feedback (dominated by supernovae), which corresponds to one supernova of $10^{51}$\,erg per 95.5\,M$_\odot$ stellar mass formed \citep[consistent with the value used in][]{Kim2020a}. Unlike EAGLE and TNG, we do not include a metallicity dependence of the energy loading \citep[see][for further details of their methods]{EAGLE1,EAGLE2,Pillepich2018}. 

Additionally, in this first work we take the metallicity of wind particles to be equal to that of the ISM cell from which each particle is launched. In reality, the hot, supernova driven wind phase is expected to have an enhanced metal fraction compared to the ISM \citep[e.g.][]{LiBryan2020, Kim2020b, Steinwandel2024b}. In Appendix \ref{Section:MetalAppendix} we perform a numerical experiment in which we increase the metallicity of launched wind particles by a constant (and somewhat arbitrary) factor of 5 and look at its effect on galaxy metallicities. In that Appendix, we show how this metal enhancement can change the normalisation of the mass-metallicity relation in \ark.  However, such a constant factor is not fully consistent with the model; unfortunately, simulating metal outflows as sourced directly by the SNe instead of being proportional to the ISM metallicity is technically challenging in our current implementation and we therefore postpone a detailed study of metallicity to future work.

In \ark, wind particles are created at a higher mass resolution (i.e.~lower mass) than the base resolution of the simulation. This allows for better sampling of the energy injection rate in high specific energy winds, as well as forming part of the cosmological \textsc{Arkenstone-Hot} refinement scheme described below. A key result of \citetalias{Smith2024a} is that to accurately follow the propagation of a high specific energy wind, a simulation must have sufficient spatial resolution through the sonic point near the base of the wind (defined to be where the outflow velocity passes the radially decreasing sound speed as it adiabatically expands), where material is rapidly accelerated away from the galaxy. \ark\ alleviates some of the difficulties in simulating high specific energy winds by ensuring the spatial resolution in the hot, low density phase of the winds remains a small fraction of the distance to the galaxy, as discussed in Sections~\ref{Section:Intro} and \ref{Section:Refinement}, thus ensuring we accurately capture the wind's acceleration. 

In this work, as in \citetalias{Smith2024a}, all \ark\ wind particles have a mass $100$ times smaller than the base resolution of the simulation (and of the wind particles in the TNG model). During extensive testing we have found a slight dependence of quantities like galaxy stellar masses with the value of this ratio in cosmological simulations, which was not present in the isolated galaxies of \citetalias{Smith2024a}. This lack of convergence is potentially a consequence of enhanced cooling at higher resolution, and is something we are investigating further. We note, however, that the changes due to this are much smaller than those arising from changes in wind loading quantities, so for this work we keep all simulations with a ratio of wind resolution to base resolution of 100. 

After launch, wind particles travel (without interacting hydrodynamically) until the cell they are in has a number density below 10 per cent of the star formation density threshold - $0.1 n_{\rm H,0} \approx 10^{-2}\,\mathrm{cm}^{-3}$ \citep[as in][]{SpringelHernquist2003,Smith2024a}. We note this is slightly higher than TNG's threshold of 5 per cent of $n_{\rm H,0}$. At this point, the wind particle returns its conserved quantities to the host cell via one of the recoupling methods described in the following section.

For low energy loading values, the corresponding low launch velocities can sometimes fall below a galaxy's escape velocity, meaning wind particles are not guaranteed to escape the galaxy. This typically happens in our variable energy loading runs, in the largest haloes in the box. These haloes are regulated by BH feedback anyway -- the failure of a supernovae driven wind to escape the galaxy is unsurprising. To prevent wind particles being permanently decoupled, we set a maximum time that a wind particle can exist without recoupling of 100\,Myr. If this timer is triggered we return this mass to the ISM, and do not inject the tracer described below to avoid refining star forming material. While the particular choice of a timer of 100\,Myr is somewhat arbitrary, we have verified that instead using timer values of 20 and 1000\,Myr have no appreciable effect on our results.

\subsection{Recoupling and refinement} \label{Section:Refinement}

\begin{table*}
    \centering
    \caption{List of simulation runs presented in this paper, with their wind mass loading, energy loading, and launch velocity kick. Halo mass scalings for variable loadings runs are calculated at $z=0$; the \ark\ scalings with $\sigma_\mathrm{DM,1D}$ apply at all redshifts. The scalings with halo mass for TNG and the variable energy loading \ark\ runs are inferred and will actually be scattered about this relation, as both actually scale with the locally measured DM velocity dispersion (and metallicity in the case of TNG).}
    \begin{tabular}{|c|c|c|c|}
        \hline
        Run & $\eta_\mathrm{M}$ & $\eta_\mathrm{E}$ & $\Delta v_{\mathrm{launch}}$ [km\,s$^{-1}$]\\ \hline
        TNG & $10.2 \left(\frac{M_{\rm h}}{10^{11}\,\mathrm{M}_\odot}\right)^{-5/6}$ & (see Fig.~\ref{fig:loadings}) & max\,$\left[350,378\left(\frac{M_{\rm h}}{10^{11}\,\mathrm{M}_\odot}\right)^{1/3}\right]$ \\ \hline
        Fixed $\eta_{\rm E}$ values,  
        & 0.3 & 0.1 & 354 \\ Constant $\eta_{\rm M}$,
        & 0.3 & 0.3 & 613 \\ (Oranges)
        & 0.3 & 0.9 & 1061 \\ \hline
        Fixed $\eta_{\rm M}$ values,  
        & 0.1 & 0.3 & 1061 \\ Constant $\eta_{\rm E}$,
        & 0.3 & 0.3 & 613 \\ (Pink/Purple)
        & 0.9 & 0.3 & 354 \\ \hline
        Varying $\eta_{\rm E}$ values,  
        & 0.3 & $0.4309 \left(\frac{\sigma_{\rm DM,1D}}{\sigma_{\rm DM,1D,0}} \right)^{-1/2} \propto M_{\rm h}^{-1/6}$ & $734\left(\frac{\sigma_{\rm DM,1D}}{\sigma_{\rm DM,1D,0}}\right)^{-1/4} \propto M_{\rm h}^{-1/12}$ \\ Constant $\eta_{\rm M}$,
        & 0.3 & $0.4309 \left(\frac{\sigma_{\rm DM,1D}}{\sigma_{\rm DM,1D,0}} \right)^{-1\,\,\,\,\,\,} \propto M_{\rm h}^{-1/3}$ & $734\left(\frac{\sigma_{\rm DM,1D}}{\sigma_{\rm DM,1D,0}}\right)^{-1/2} \propto M_{\rm h}^{-1/6\,\,\,}$ \\ (Blues)
        & 0.3 & $0.4309 \left(\frac{\sigma_{\rm DM,1D}}{\sigma_{\rm DM,1D,0}} \right)^{-3/2} \propto M_{\rm h}^{-1/2}$ & $734\left(\frac{\sigma_{\rm DM,1D}}{\sigma_{\rm DM,1D,0}}\right)^{-3/4} \propto M_{\rm h}^{-1/4\,\,\,}$ \\ \hline
    \end{tabular}
    \label{tab:sim_runs}
\end{table*}

A key component of the \textsc{Arkenstone-Hot} model described in \citetalias{Smith2024a} is the introduction of the displacement recoupling method. When the host cell of a wind particle meets the density criterion described in Section~\ref{Section:ArkenstoneWinds}, the wind particle injects its mass, momentum, energy, and metals into that cell. However, if a high resolution, high specific energy wind particle were to inject its energy immediately into a more massive host cell, it would be diluted, leading to significant numerical overcooling. To prevent dilution of the high specific energy of a wind particle, we first displace the existing contents of the host cell to its neighbouring cells before the wind particle recouples. Importantly, this also maintains the existing regular mesh structure so the hydrodynamics scheme continues to work correctly, which would not be the case if the standard \textsc{Arepo} refinement was used. 

When recoupling occurs, a passive scalar ``dye" is injected, which is then advected as a tracer with the hydrodynamics scheme as a conserved quantity. Cells are flagged as ``hot wind cells" if the ratio of this quantity to the cell mass, $f_\mathrm{w}$, is greater than a threshold value $f_\mathrm{w,thresh}$. In this work this threshold is set as $f_\mathrm{w,thresh} = 1000$, the choice of which is described below. Displacement recoupling only occurs for host cells that do not already have a high temperature, or those already flagged as part of the hot wind. In \citetalias{Smith2024a}, an additional requirement for cells to maintain a higher pressure than their surroundings (by retaining more mass during the displacement process) was used. However, we found in cosmological simulations that this occurred too frequently and had the effect of diluting our wind energy, so in this work we do not use this criterion.\footnote{Additional criteria can also apply to prevent a cell from displacing material to its neighbour \citepalias[see][for details]{Smith2024a}, though these occur rarely.} If a displacement recoupling does not happen a ``simple'' recoupling occurs instead, whereby wind particle quantities are directly injected into its host cell, as happens in TNG. 

Once displacement occurs, the host cell has a lower mass. After recoupling, we wish to prevent this cell from derefining back to the base resolution of the simulation, as high spatial resolution around the sonic point is required to correctly resolve the propagation of a wind \citep{Smith2024a}. We therefore maintain the target mass of hot wind cells at the same mass as the wind particles themselves. This is the refinement scheme that allows \textsc{Arkenstone} to successfully resolve the driving of high specific energy winds from close to the galaxy. 

However, we do not want to maintain higher resolution in the simulation indefinitely and refine the entire CGM, as this becomes prohibitively expensive. This requires a method of relaxing the refinement as the wind gets further from the galaxy. The fiducial \ark\ implementation was designed and tested in isolated galaxy simulations that had an easily defined frame of reference, making it simple to implement a radial (i.e. galactocentric distance) criterion to prevent lower mass cells from de-refining. However, for the use of the \ark\ model in cosmological simulations, a different method of managing the enhanced resolution is needed, as each galaxy has its own frame of reference. The simplest option could be to use a fixed refinement radius for all galaxies, though because haloes have different sizes this would mean maintaining high resolution for large distances around low mass galaxies, to ensure the same radius also encompasses the sonic point of massive haloes. To do this more efficiently for haloes of different masses therefore requires the wind particles to have knowledge of the halo in which they are created.

Our new refinement scheme still makes use of the passive scalar dye injected by wind particles, described earlier, but instead the value decays over time. The scheme is designed to resolve the base of the galactic wind around galaxies of different masses, without refining the entire CGM. To do this, we conservatively aim to maintain high resolution in the wind until $0.2 R_\mathrm{200}$\footnote{We define $R_\mathrm{200}$ as the radius within which the mean density is 200 times the critical density of the universe at a given redshift.} around a given galaxy. The particular choice of $0.2 R_\mathrm{200}$ is somewhat arbitrary, as there is no physical reason for maintaining the resolution to this distance, but the motivations for it are two-fold. Firstly, we cannot easily predict the location of the sonic point around a given galaxy in a cosmological simulation, as it is affected by the galaxy's geometry and environment. Secondly, the virial radius scales with halo mass, so that we avoid the issues raised by using a fixed radius. Additionally, an estimate of the virial radius is available from local properties of gas cells that launch wind particles, as described below. 

In post-processing of an initial simulation, we verified that the radius of $0.2 R_\mathrm{200}$ comfortably encompasses the sonic point of the hot wind across a wide range of galaxy masses. There may be more efficient ways of determining a refinement radius that we will investigate in future work, but the current scheme works well empirically and so is used throughout this paper.

After wind particles recouple and the dye is injected, we implement an exponential decay of the scalar dye with time,
\begin{equation}
    f_\mathrm{w}(t + \delta t) = f_\mathrm{w}(t) e^{-\frac{\delta t}{t_0}}, 
\end{equation}
where $t_0$ is a decay constant we set to a fixed $t_0 = 10$\,Myr. This therefore requires us to change the initial value of the dye, so that by the time the wind reaches $\sim0.2R_\mathrm{200}$ the scalar tracer falls below a value of $f_{\rm w, thresh} = 1000$ and cells are no longer flagged. We emphasise that the specific values of $t_0$ and $f_{\rm w, thresh}$ do not matter, it is the combination of them with the current value of the dye that determines a time until a cell begins to derefine. Though we note that using high values of $f_{\rm w, thresh}$ (together with high values of $f_{\rm w}$ at recoupling; see below) is preferred so that mixing with pristine, unflagged gas (which can also decrease the dye value) does not significantly affect the timer.

To estimate $R_\mathrm{200}$ from local properties at launch, we use the 1D dark matter velocity dispersion, which has been found to be correlated with the maximum circular velocity of a halo \citep[$v_\mathrm{max} \simeq 1.45 \sigma_\mathrm{DM, 1D}$,][]{Okamoto2010}, and been used in a number of simulations for the launching of winds themselves \citep[e.g.][]{Oppenheimer2008,Illustris1}. This is calculated in a kernel-weighted fashion over the nearest 64 dark matter particles to the host gas cell. It can be shown for a Navarro-Frenk-White (NFW) halo \citep{Navarro1997} that the maximum circular velocity $v_\mathrm{max}$ can be linked to the virial velocity $v_\mathrm{200}$ via 
\begin{equation} \label{Eqn: Vmax}
    v_\mathrm{max} \simeq 0.465 \sqrt{f(c)} \ v_\mathrm{200},
\end{equation}
where $c$ is the concentration parameter of the NFW halo and 
\begin{equation}
    f(c) = \frac{c}{\ln (1 + c) - c/(1+c)}.
\end{equation}
$R_\mathrm{200}$ is then linked to the virial velocity via
\begin{equation} \label{Eqn: Vvir}
    v_\mathrm{200} = \sqrt{\frac{G M_\mathrm{200}}{R_\mathrm{200}}} = \sqrt{\frac{4 \pi G}{3} \Delta_\mathrm{c} \rho_\mathrm{crit}(z)} \ R_\mathrm{200},
\end{equation}
where $\Delta_\mathrm{c} = 200$ is the overdensity within $R_\mathrm{200}$, and $\rho_\mathrm{crit}(z) = 3 H(z)^2 / 8 \pi G$ is the critical density at the redshift of interest. 

Combining equations (\ref{Eqn: Vmax}) and (\ref{Eqn: Vvir}) with the result of \citet{Okamoto2010} leads to the following relation:
\begin{equation}
    R_\mathrm{200} \simeq 3.118 \ \sqrt{\frac{3}{800 \pi G \rho_\mathrm{crit}(z) f(c)}} \ \sigma_\mathrm{DM,1D}. 
\end{equation}
During the simulation, we will not know the concentration parameter $c$ for a given halo a priori, so we currently use a fixed value of $c=2$, which minimises the function $f(c)$ and so provides the most conservative estimate of $R_\mathrm{200}$. 

From the launch quantities defined in Equations (\ref{Eqn: LaunchV}) and (\ref{Eqn: LaunchU}), the Bernoulli velocity is given by 
\begin{equation}
    v_\mathrm{B} = \sqrt{(\Delta v_\mathrm{w})^2 + 2 \gamma u_\mathrm{w}},
\end{equation}
which is the characteristic velocity for a wind launched with these properties, and $\gamma = 5/3$ is the adiabatic index of the gas. In reality, due to a number of factors, the wind will not actually be at this velocity, and so we introduce an additional factor $k=0.25$ to encompass such uncertainties. Again, we emphasise that the specific values entering the calculation do not matter, only the approximate scaling, and we have empirically found this to work.

The initial tracer value given to the wind particles to pass on to their host cell when recoupling occurs is therefore set to be
\begin{equation}
    f_\mathrm{w, init} = f_\mathrm{w, thresh} \exp{\frac{0.2 R_\mathrm{200}}{k v_\mathrm{B} t_0}}.
\end{equation}
Gas thus travels approximately $0.2 R_{200}$ before the tracer decays below $f_{\rm w}$, so cells maintain a higher resolution within that region. If the value of the tracer carried by a recoupling wind particle is higher than the existing tracer in its host cell, the tracer is overwritten. 

Because the mass of wind particles and the cells they recouple into are 100 times smaller than the target cell mass of the base simulation, immediately derefining cells to the base target mass would result in an abrupt jump in mass resolution. To avoid this, we allow a more gradual change in resolution by continuing the decay of the tracer past $f_\mathrm{w, thresh}$ until the tracer reaches a new threshold value, which we set in this work to be $f_\mathrm{w, ref} = 0.1$. During this time we link the target cell mass directly to the tracer value via 
\begin{equation}
    m_\mathrm{target} = \frac{m_0}{\sqrt{f_\mathrm{w} / f_\mathrm{w, ref}}},
\end{equation}
so that the resolution smoothly degrades until the cell mass reaches the base resolution, $m_0$, of the simulation.

\subsection{Mass and energy loadings} \label{Section:Loadings}

\begin{figure}
    \centering
    \includegraphics[width=0.99\linewidth]{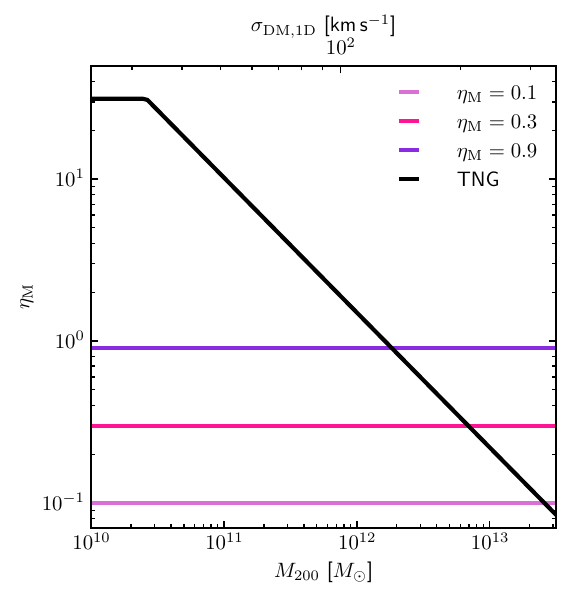}
    
    \includegraphics[width=0.99\linewidth]{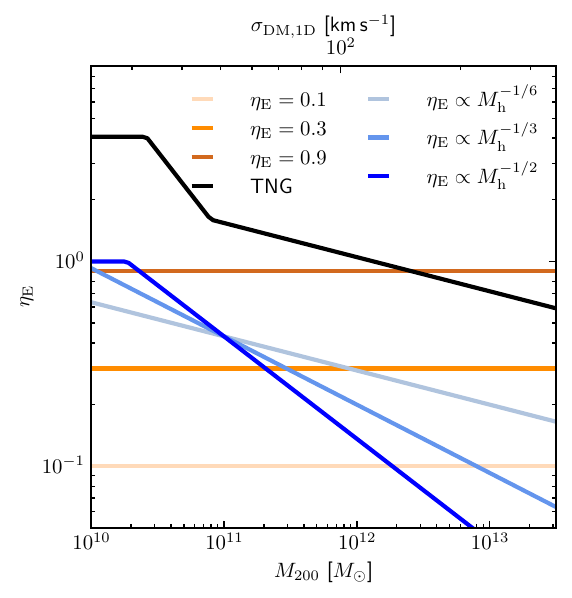}
    \caption{Mass (top panel) and energy (bottom panel) loadings at $z=0$ for all \ark\ simulations in this paper, in comparison to scalings from TNG \citep[from the $z=0$ fits to $\eta_{\rm M}$ and $v_{\rm w}$ in][]{Pillepich2018}. Both mass and energy loadings in \ark\ runs are generally much lower than in TNG. Throughout most plots in this paper, shades of orange show simulations with constant $\eta_\mathrm{E}$ values and fixed $\eta_\mathrm{M}=0.3$; pink/purple lines show simulations with constant $\eta_\mathrm{M}$ values and fixed $\eta_\mathrm{E}=0.3$; shades of blue show simulations with variable energy loading with a fixed $\eta_\mathrm{M}=0.3$. We note that the input loading values actually applied in the simulations for the latter and TNG have an associated scatter, due to the use of the locally measured DM velocity dispersion (and metallicity in the case of TNG) in their calculation. \ark\ runs with variable energy loadings are capped at an $\eta_\mathrm{E} = 1$.} 
    \label{fig:loadings}
\end{figure}

In this paper we present results from a total of 10 cosmological simulations. We compare the predictions from the \ark\ model to the output of the existing TNG model (but without MHD), which we ran using the same initial conditions and simulation configuration. Specifically, we have three runs with a fixed mass loading $\eta_\mathrm{M} = 0.3$, and fixed energy loadings of $\eta_\mathrm{E} = 0.1$, 0.3, and 0.9 (shown in all plots in shades of orange). This is complemented with three runs with a fixed energy loading $\eta_\mathrm{E} = 0.3$, and fixed mass loadings of $\eta_\mathrm{M} = 0.1$, 0.3, and 0.9 (shown in pink/purple).

Our remaining runs (shown in shades of blue) have a variable energy loading. As discussed in e.g. \citet{Carr2023} and later in this paper, variations in the mass loading of the high specific energy wind do not have as large an impact as variations in energy loading. For these first numerical experiments, we therefore focus only on variable \textit{energy} loadings. We scale these loadings with the 1D dark matter velocity dispersion, which, as described in the previous section, is linked to the halo mass. These are motivated by findings from previous simulations and semi-analytic models suggesting higher energy loadings are required to regulate low mass galaxies \citep[e.g.][]{Pandya2021,Carr2023,Voit2024b}. We characterise this variable energy loading as 
\begin{equation} \label{Eqn:VaretaE}
    \eta_\mathrm{E} = \eta_\mathrm{E,norm} \left(\frac{\sigma_{\rm DM,1D}}{\sigma_{\rm DM,1D,0}} \right)^{3\alpha}.
\end{equation}
The slope $\alpha$ is the expected scaling we wish to realise with halo mass; the additional factor of $3$ comes due to the relationship $\sigma_{\rm DM,1D} \propto M_{\rm h}^{1/3}$ (at fixed redshift). $\sigma_{\rm DM,1D,0}$ is the characteristic DM velocity dispersion for a $10^{11}$\,M$_\odot$ halo at $z=0$, which we take to be $51$\,km\,s$^{-1}$. At this halo mass we then take the expected energy loading to be $\eta_\mathrm{E,norm} = 0.4309$ (equivalent to a normalisation of 0.2 for a halo mass of $10^{12}$\,M$_\odot$ at $z=0$ when $\alpha=-1/3$). Equation (\ref{Eqn:VaretaE}) is the input energy loading for these variable energy loading simulations, which then approximately follows the halo mass scaling
\begin{equation}
    \eta_\mathrm{E} \approx \eta_\mathrm{E,norm} \left[\frac{M_{\rm h}}{10^{11}\,\mathrm{M}_\odot} 
    \left( \frac{H(z)} {H_0}\right)^{1/3}
    \right]^{\alpha}. 
    \label{eq:eta_e_scaling}
\end{equation}
Our three runs forming the final group still include a fixed mass loading $\eta_\mathrm{M} = 0.3$, but include this parametrisation of a variable energy loading with slopes $\alpha = -1/6, -1/3,$ and $-1/2$. For simplicity, we generally drop the redshift dependence in Eq.~\ref{eq:eta_e_scaling} when discussing runs with this scaling.
All runs in this paper and their basic properties are summarised in Table~\ref{tab:sim_runs}. 

We additionally cap the energy loading in all \ark\ simulations at $\eta_\mathrm{E,max} = 1$. Realistically, the actual amount of energy available to drive winds out of the galaxy will be less than $\eta_\mathrm{E,max}$, as work will need to be done both to inflate bubbles and drive turbulence in the ISM. An energy loading $\eta_\mathrm{E} > 1$ is clearly too high, which is why we cap our runs at $\eta_\mathrm{E,max}$, but to better understand the true amount of energy available to drive winds we must turn to high resolution, more idealised simulations \citep[e.g. TIGRESS,][]{KimOstriker2018}, loadings from which we will investigate in future work.

We show our loadings for both mass (top panel) and energy (bottom panel) in Fig.~\ref{fig:loadings}, compared to the input values for TNG \citep[from the fits of][]{Pillepich2018}. In TNG, the energy loading (dependent on metallicity) and launch velocity (dependent on redshift and with a minimum value of $350\,{\rm km\,s^{-1}}$) are model inputs, and this then sets the mass loading. The values for our variable energy loading simulations and for TNG actually show a scatter around the analytic lines shown in Fig.~\ref{fig:loadings}, induced by scatter in the DM velocity dispersion (and due to the metallicity dependence, in the case of TNG). It is immediately obvious from the top panel that the mass loadings of all \ark\ runs are dramatically smaller than for TNG in low mass galaxies -- the central mass loading value used in most \ark\ runs is 5, 34, and even 100 times lower than TNG for haloes of $10^{12}$, $10^{11}$, and $10^{10}$\,M$_\odot$, respectively.  

The bottom panel of Fig.~\ref{fig:loadings} shows that \textit{energy} loadings are also high in TNG, as a direct consequence of the high mass loadings and fixed thermal energy fraction of the winds. Even the maximum energy loading in our variable runs, which are capped at $\eta_\mathrm{E} = 1$, still have energy loadings at least 3.6 times lower than TNG for low mass galaxies. 

We re-emphasise that while both our mass and energy loadings are generally lower than in TNG, an important change is that the \textit{specific} energy loading, $\eta_{\rm E} / \eta_{\rm M}$, is higher. The rest of this paper is dedicated to exploring the implications of this. Drastically reduced amounts of material are thrown out of the ISM in \ark\ simulations compared to TNG. If this is not offset by the prevention of gas joining the ISM in the first-place, galaxies will grow far too massive. As we will see, however, with high specific energy winds and appropriately scaled energy loadings, this does not occur.

\begin{figure*}
    \centering
    \includegraphics[width=0.9\linewidth]{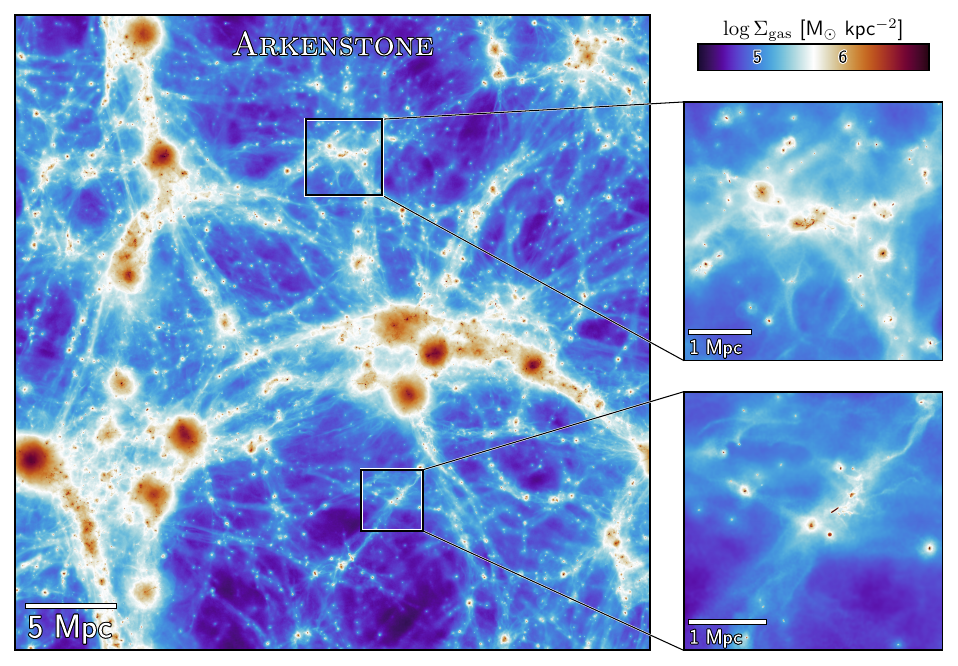}
    \includegraphics[width=0.9\linewidth]{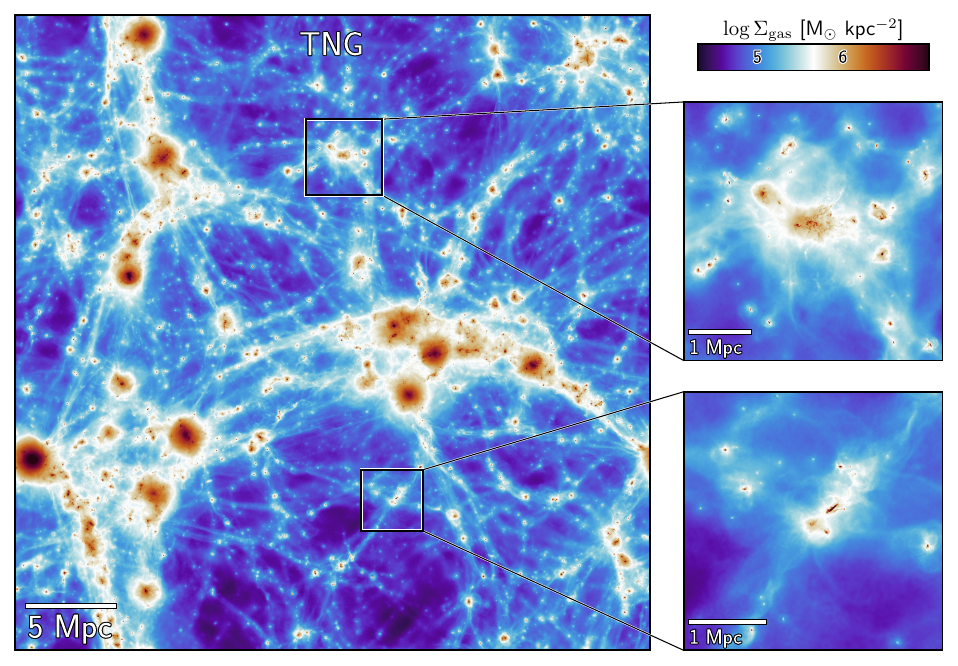}
    \caption{Gas surface density maps of the entire simulation box at $z=0$ for the \ark\ variable energy loading run with $\eta_{\rm E} \propto M_{\rm h}^{-1/3}$ (top panels) and TNG (bottom panels). Zoomed regions highlight two haloes that show particularly different density structures in the two simulations. Gas densities near to galaxies are decreased with \ark, with more gas pushed to large distances.}
    \label{Fig:DensMap}
\end{figure*}

\begin{figure*}
    \centering
    \includegraphics[width=0.9\linewidth]{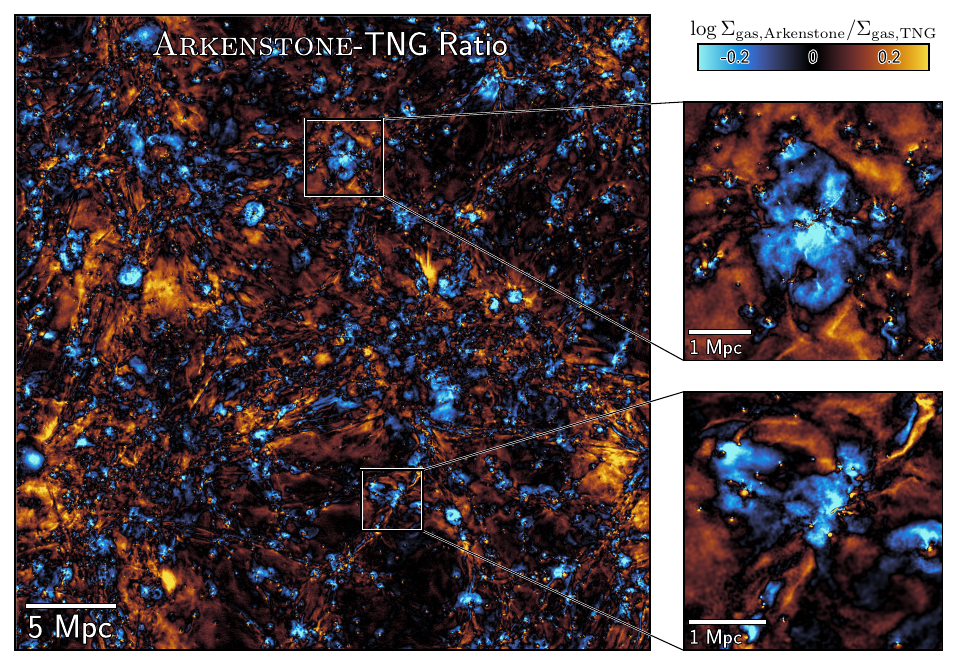}
    \caption{Ratio of the surface density maps shown in Fig.~\ref{Fig:DensMap}, highlighting the differences between the runs. \ark\ has considerably more gas (shown in orange) in the outskirts of haloes and into the IGM, due to winds extending to large distances from central galaxies. Interestingly, within that the CGM sometimes appears somewhat depleted in \ark\ compared to TNG (shown in blue).}
    \label{Fig:DensRatioMap}
\end{figure*}

\section{Results} \label{Section:Results}

In this section we present our simulations results from all cosmological simulations, and as a comparison we show results from \ark\ alongside those from the existing TNG model (which is calibrated to match certain observations). Where available, we also show a comparison with physical quantities estimated from observational data for selected quantities. We note, however, that in this work we do not forward model our simulation predictions into the observational plane, so the comparisons with observational data are intended to be qualitative and should be treated with caution.

\subsection{Large-scale gas distribution} \label{Section:FirstLook}

In Fig.~\ref{Fig:DensMap} we show gas surface density maps for the \ark\ simulation with $\eta_{\rm E} \propto M_{\rm h}^{-1/3}$ (top panels) and TNG (bottom panels). Orange areas highlight the high gas densities in the centre of haloes, dark blue regions indicate voids. With \ark, gas densities close to galaxies are reduced, with a corresponding increase at large radii. This is particularly noticeable in the two haloes highlighted in the zoomed panels on the right, where the dense gas present in TNG is reduced in the \ark\ run.

The differences between the simulations are emphasised in Fig.~\ref{Fig:DensRatioMap}, which shows the ratio of the two maps in Fig.~\ref{Fig:DensMap}. Here, blue regions highlight where the gas surface density is lower in \ark\ than TNG, which occurs close to many galaxies in the simulation volume. Orange regions show where gas surface densities are higher in the \ark\ run, which generally happens on the outskirts of haloes and into the IGM. We note that many of the changes seen in gas density here materialise well outside the virial radii of haloes, affecting the outer CGM and IGM, showing how high specific energy winds can have important impacts far from the galaxies that launch them. Detailed properties of the IGM in \ark\ simulations will be studied in future work, although we note that a generic prediction is a higher IGM gas fraction, potentially in agreement with recent observations \citep{Connor2025}.

\subsection{Galaxy properties at $z=0$} \label{Section: z0results}

\begin{figure}
    \centering
    \includegraphics[width=0.8\linewidth]{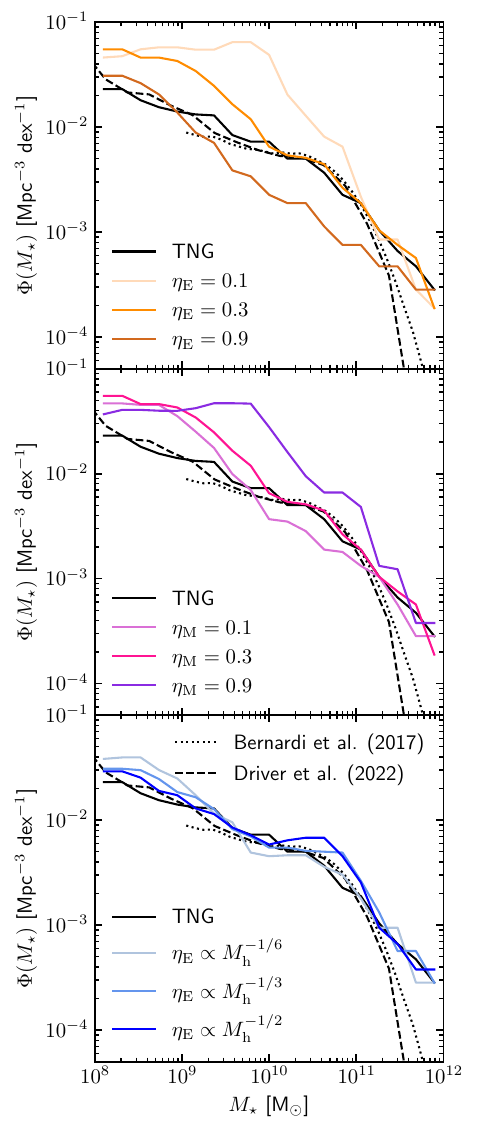}
    \caption{Stellar mass functions at $z=0$ for simulations with: 1) constant energy loadings and fixed mass loading of $\eta_{\rm M} = 0.3$ (top panel), 2) constant mass loadings and fixed energy loading of $\eta_{\rm E} = 0.3$ (middle panel), and 3) simulations with variable energy loading with different slopes $\alpha$ and a fixed mass loading of $\eta_{\rm M} = 0.3$ (bottom panel - we remind the reader that we have dropped the redshift dependence on $\eta_{\rm E}$ in this panel for convenience). Each panel also shows the result from TNG (black solid line) and observational data from  \citet{Bernardi2017} (black dotted line) and \citet{Driver2022} (black dashed line). Lines shown are the median of the data, smoothed over the nearest two bins. For the hot, light winds modelled by \ark, energy loading affects the normalisation and slope of the SMF much more than mass loading.}
    \label{fig:z0_SMF}
\end{figure}

\begin{figure}
    \centering
    \includegraphics[width=0.8\linewidth]{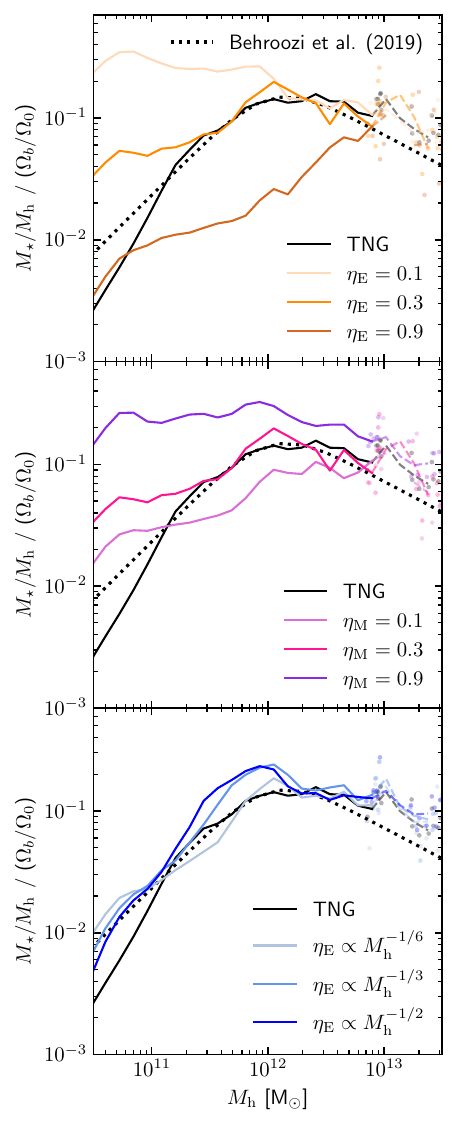}
    \caption{Ratios of stellar mass to halo mass at $z=0$ as a function of halo mass, normalised by the universal baryon fraction \citep[$f_\mathrm{B} = 0.16$,][]{Planck2015Parameters}. Median lines have the same styles as Fig.~\ref{fig:z0_SMF}, with comparison data from \citet{Behroozi2019}. For the largest haloes in our box we plot individual scatter points, and change our median lines to be dashed. This is to emphasise that results at high masses should be treated with caution, due to the lack of massive haloes and the dominance of the TNG BH model here. Fixed energy loadings tend to overproduce stars at the low mass end, which can be alleviated by introducing a variable energy loading.}
    \label{fig:z0_mstar_mh}
\end{figure}

From this Section to Section~\ref{Section:RedshiftEvo}, we show a number of population-level results for the properties of galaxies and their gaseous haloes in our simulated boxes. For all plots except the SFR density and stellar mass function, solid lines show the median of each quantity, smoothed over the nearest two points for clarity. 

In all panels, the solid black line shows the quantity for the TNG model run with the same initial conditions and with the same simulation configuration. For \ark\ runs, lines with shades of orange show the results for runs with a fixed \textit{mass} loading ($\eta_\mathrm{M} = 0.3$) and several constant energy loading values. Shown in pink and purple are runs with a fixed \textit{energy} loading ($\eta_\mathrm{E} = 0.3$) and several constant mass loading values. Finally, lines in shades of blue show results for runs with a variable energy loading. Other lines in each panel show observational constraints where available, though we note that a more detailed comparison with observations is deferred to future work. 

We further note that due to our limited box size of ($36.9\,$Mpc)$^3$, our results at high stellar and halo masses are limited by poor statistics and large cosmic variance for large objects. For this reason, and the uncertain interplay between \ark\ and the TNG BH model, we suggest caution in interpreting results at these masses. We show individual data points for the largest few haloes in the box (above $M_{\rm h} \sim 10^{13}\,$M$_\odot$), and plot the medians as dashed lines.

\subsubsection{Stellar mass functions}

In Fig.~\ref{fig:z0_SMF}, we show the SMFs at $z=0$ for all of our runs. The solid black line in all panels shows the SMF for the TNG model, the parameters of which were tuned to reproduce this quantity (although using different observational data). In the top panel, showing effects of changing energy loading while holding mass loading constant, the cumulative feedback effect of energy loading of a hot wind is very clear. The palest line shows the lowest energy loading, $\eta_\mathrm{E}=0.1$, which is evidently ineffective at regulating star formation across a wide range of galaxy masses. The intermediate energy loading of $\eta_\mathrm{E} = 0.3$ is shown in the middle orange colour, which brings the SMF down closer to the observations, especially for galaxies at a stellar mass of $\sim10^{10}$\,M$_\odot$. This run still shows an excess of galaxies with $M_\star < 10^9$\,M$_\odot$, suggesting a fixed energy loading of 0.3 is not enough to regulate these dwarf galaxies. We discuss this further below. The darkest line in the top panel shows the highest energy loading, $\eta_\mathrm{E} = 0.9$, which overregulates star formation in galaxies with $M_\star > 10^9$\,M$_\odot$, but produces a distribution closer to the observations for $M_\star < 10^9$\,M$_\odot$. This points to the need for a higher energy loading in lower mass systems.

The middle panel of Fig.~\ref{fig:z0_SMF} shows simulations with changes in mass loading for a fixed energy loading of $\eta_\mathrm{E} = 0.3$. The effect of these changes in mass loading is smaller than the changes from varying the energy loading, though the amplitude of the SMF does decrease slightly with decreasing mass loading (i.e. less mass ejection results in less star formation). As we discuss later, this is due to a lower mass loading leading to a hotter, faster wind, resulting in a higher level of preventative feedback due to CGM heating. This is the opposite behaviour to regulation via ejective feedback, where increasing the mass loading tends to lower the star formation by reducing the amount of material in the ISM (as long as there is still enough energy to throw the mass far enough). The run with $\eta_{\rm M} = 0.9$ (and $\eta_E=0.3$) failed to regulate star formation at low redshifts across a wide halo mass range -- this wind has too low a specific energy to provide effective preventative feedback while also having too small a mass loading to act as an efficient source of ejective feedback.

Our runs with an energy loading scaling with halo mass (via the dark matter velocity dispersion) are shown in the bottom panel of Fig.~\ref{fig:z0_SMF}. The dependence we have implemented leads to a higher energy loading at lower galaxy masses, which brings the SMF into much better agreement with observations. The slope of the variable energy loading makes a slight difference to the amplitude of the SMF for low mass galaxies, with the steepest value appearing to give the closest match to the observational data.

\subsubsection{Stellar mass -- halo mass relation}

A similar picture emerges in the SMHM relation in Fig.~\ref{fig:z0_mstar_mh}, here normalised by the cosmic baryon fraction. Recall that this relation (although different data) was used to calibrate the TNG model. The different (constant) values of energy loading in the top panel show the strong sensitivity of the SMHM to the value of $\eta_{\rm E}$, with the gradient getting steeper with higher energy loading. The lowest energy loading clearly leads to a significant overproduction of stars at the low-mass end, whereas the high-mass end closely follows TNG, likely due to the dominance of the AGN feedback model. An energy loading of 0.3 reduces the stellar mass across all halo masses, most notably below $10^{12}$\,M$_\odot$, though there is still an overproduction of stars in the dwarf regime. The highest energy loading of 0.9 dramatically and too effectively reduces the stellar mass of many galaxies -- the SMHM relation lies nearly an order of magnitude below the observational data at a Milky Way halo mass of $\sim\!10^{12}$\,M$_\odot$. In the dwarf regime the SMF is closer to the observations, again pointing towards the need for higher energy loading in low mass galaxies. Notably, the higher energy loadings also lead to somewhat reduced stellar masses in the largest haloes in our box - this is most likely due to strong winds affecting the progenitors of these objects at early times.

The mass loading of high specific energy winds can also affect the number of stars formed in lower mass haloes, but much less significantly than energy loading (we note that the gradient of this relation is strongly dependent on energy loading). Lower mass loadings, via hotter, faster winds, lead to fewer stars forming in lower mass haloes. We note that this difference gets smaller at halo masses above $\sim\!10^{12.5}\,{\rm M}_\odot$, though as we have discussed, the interpretation here is complicated by the AGN feedback. 

As with the SMF, adopting a variable energy loading with halo mass brings the SMHM for \ark\ into better agreement with observations, by further decreasing the stellar masses of low mass haloes. The slope of the variable energy loading directly affects the slope of the SMHM at low masses, with the steepest relation leading to more suppression below $\sim\!10^{11}$\,M$_\odot$, and less above that. Our results seem to favour a steeper gradient of $\eta_{\rm E}$, closer to that of \citet{Carr2023} than that of \citet{Voit2024a}. At the high mass end, the energy loading becomes small, and so the SMHM relation is set primarily by the TNG BH model, though of course the early evolution of such objects may also be different. 

\subsubsection{Stellar mass -- black hole mass relation}
\begin{figure}
    \centering
    \includegraphics[width=0.8\linewidth]{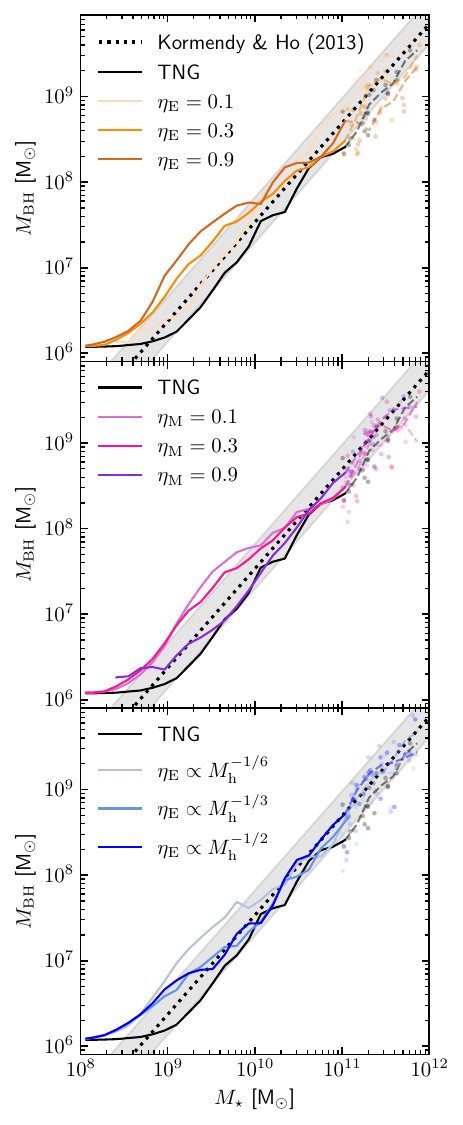}
    \caption{BH-stellar mass relation at $z=0$. Line and points are displayed in the same way as Fig.~\ref{fig:z0_mstar_mh}. A fit to observational data from \citet{KormendyHo2013} is shown as a dotted line, with the grey shaded region showing the 0.29 dex intrinsic scatter about this relation. BHs in \ark\ tend to be slightly more massive than in TNG by $z=0$, likely causing changes in quenching behaviour due to TNG's feedback model.}
    \label{fig:z0_mstar_mbh}
\end{figure}

Though we have repeatedly discussed the caveats associated with the complex interplay between \ark\ and the fiducial TNG BH model, particularly for galaxies at higher masses, it is still worth examining basic galaxy versus BH scaling relations for the \ark\ models. In Fig.~\ref{fig:z0_mstar_mbh} we show the stellar mass - black hole mass relation for our simulations. This was another of the observationally based relations used to calibrate the TNG model. The inclusion of MHD increases this quantity at fixed stellar mass slightly in the TNG framework, so that the full fiducial TNG model lies slightly \textit{above} the median \citet{KormendyHo2013} relation \citep[see][]{Pillepich2018}.

In most \ark\ runs, BHs are a factor of a few larger than the equivalent in TNG. Investigating the reasons \textit{why} this happens is beyond the scope of this paper, and will be studied in future work. But we note that none of the parameters for the AGN feedback model used in TNG were changed for the \ark\ runs, despite a dramatic change the stellar driven wind properties. The larger BHs \textit{do} have implications for galaxy quenching, however, due to the BH mass dependent switch that determines when TNG's powerful kinetic feedback mode is activated \citep[typically in galaxies with stellar masses above of $10^{10.5}\,$M$_\odot$,][]{Weinberger2017a}. 

There appears to be a slight dependence of the amplitude of the stellar mass -- black hole mass relation on the energy loading of the wind, with higher energy loadings leading to slightly more massive black holes in the dwarf regime, and slightly less massive black holes in the group regime. This implies that highly energy loaded winds can potentially suppress the growth of massive black holes in large haloes, though we defer a full investigation of \ark\ in group and cluster-scale haloes to future work.

\subsubsection{Galaxy sizes}
\begin{figure}
    \centering
    \includegraphics[width=0.8\linewidth]{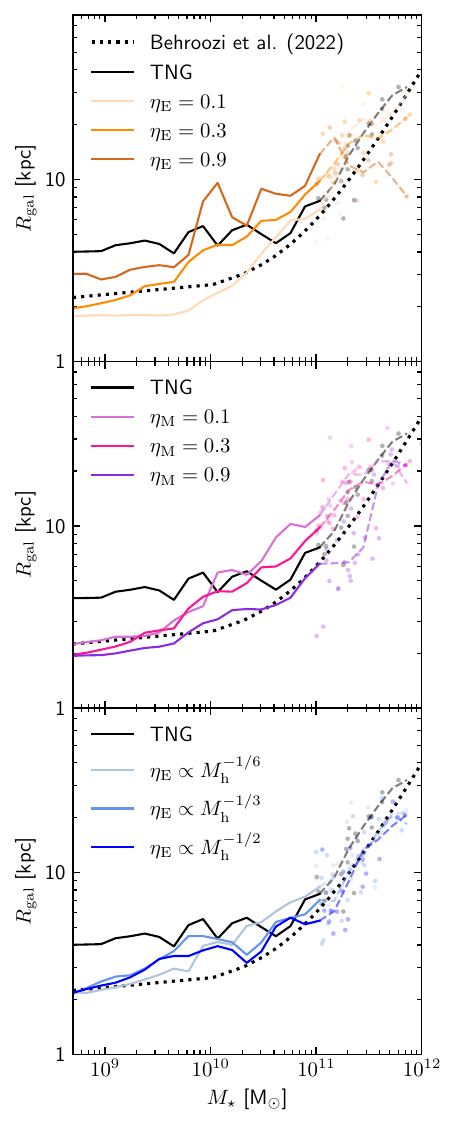}
    \caption{Galaxy sizes at $z=0$, defined as the stellar half mass radii, as a function of stellar mass. Lines and points are displayed in the same way as Fig.~\ref{fig:z0_mstar_mh}, with observational data from \citet{Behroozi2022} included as a dotted line. \ark\ tends to produce smaller low-mass galaxies, with the growth of larger galaxies happening at slightly smaller masses than in TNG.}
    \label{fig:z0_mstar_rgal}
\end{figure}

As in previous studies that find that stellar winds have a significant impact on galaxy sizes, we find that the energy loading of our galactic winds plays a significant role in the stellar sizes of galaxies at $z=0$. In Fig.~\ref{fig:z0_mstar_rgal}, we show the median 3D stellar half-mass radius as a function of galaxy stellar mass. TNG was calibrated to reproduce this relationship, though using a different observational dataset than the one presented here. The TNG runs (solid black line) slightly over-estimate the size of galaxies compared to the latest observational data \citep{Behroozi2022}, though we note that there are many caveats to such a comparison without full mock observations, and that MHD (as present in the original TNG simulations) tends to slightly reduce the size of stellar discs \citep{Pillepich2018}. It is also worth noting that the TNG model significantly improved on the over-estimation of galaxy radii found in the original Illustris simulation, pointing again to the sensitivity of galaxy sizes to the feedback implementation \citep{Genel2018}. 

With \ark, the sizes of galaxies at a given stellar mass has a notable dependence on $\eta_\mathrm{E}$, with higher energy loadings leading to larger galaxies at nearly all galaxy masses (shown in the top panel of Fig.~\ref{fig:z0_mstar_rgal}). The central value of $\eta_\mathrm{E}=0.3$ provides the best match to the observational galaxy sizes at the low mass end, though we note that the turning point where the slope steepens happens at lower masses in \ark\ compared to TNG. This could be linked to the galaxy becoming quenched by the TNG BH model's kinetic feedback, which kicks in at slightly lower masses in the \ark\ runs due to larger BH masses. We postpone a more detailed study of the structure of galaxies to future work. The bottom two panels show that galaxy sizes are less affected by mass loading, and also change little when changing the slope of a variable energy loading. The variable loadings in general lie reasonably close to the observational estimates, although again a full forward modelling of galaxy sizes would be needed to draw more concrete conclusions.

\subsection{CGM properties at $z=0$} \label{Section: z0results_CGM}

\begin{figure}
    \centering
    \includegraphics[width=0.8\linewidth]{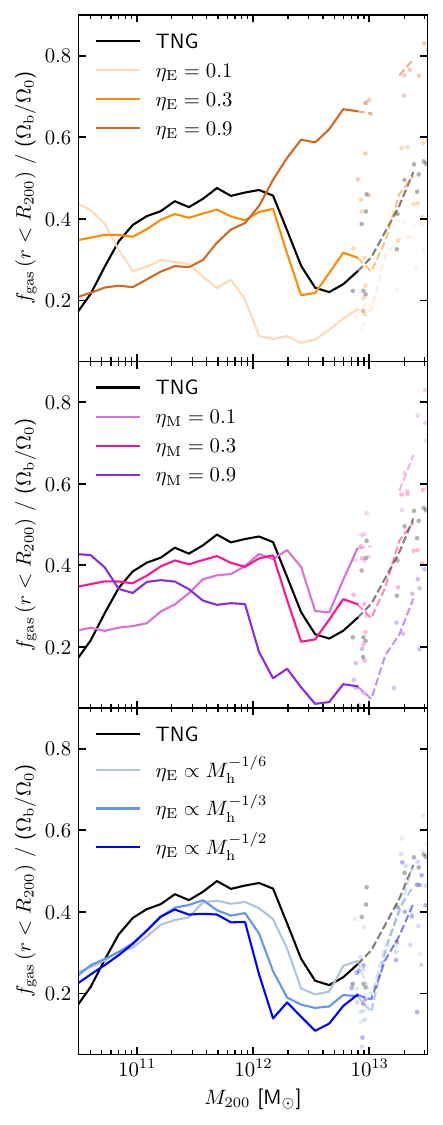}
    \caption{Gas fractions within $R_{200}$ at $z=0$, normalised by the universal baryon fraction \citep[$f_\mathrm{B} = 0.16$,][]{Planck2015Parameters}. Lines and points are displayed in the same way as Fig.~\ref{fig:z0_mstar_mh}. Observational data exists only for the high mass end, which is mostly affected by the TNG BH model, so we do not include it here. Gas fractions at $z=0$ increase gradually with halo mass for most of the \ark\ simulations, until the kinetic mode of BH feedback begins to act.}
    \label{fig:z0_mstar_fgas}
\end{figure}

\begin{figure}
    \centering
    \includegraphics[width=0.8\linewidth]{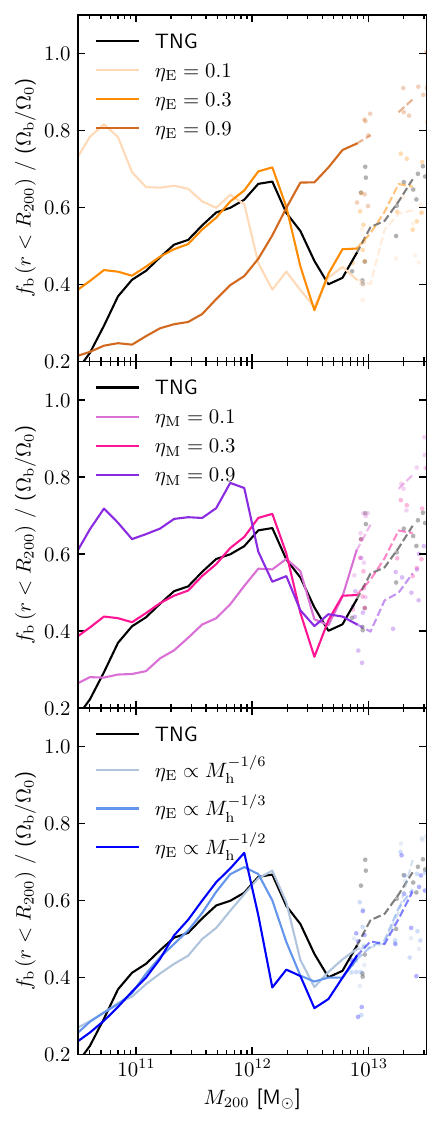}
    \caption{Baryon fractions within $R_{200}$ at $z=0$, normalised by the universal baryon fraction \citep[$f_\mathrm{B} = 0.16$,][]{Planck2015Parameters}. Lines and points are displayed in the same way as Fig.~\ref{fig:z0_mstar_mh}. Baryon fractions show a similar shape to gas fractions, increasingly gradually with halo mass until the kinetic mode of BH feedback begins to act.}
    \label{fig:z0_mstar_fbary}
\end{figure}

\begin{figure}
    \centering
    \includegraphics[width=0.8\linewidth]{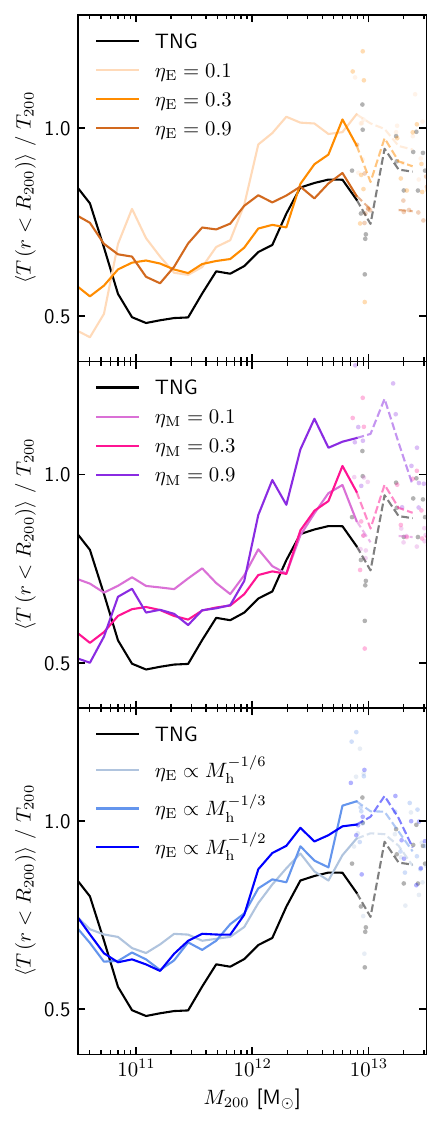}
    \caption{Mass-weighted mean circumgalactic medium temperatures within $R_{200}$ at $z=0$, normalised by the nominal virial temperature $T_{200}$. Lines and points are displayed in the same way as in Fig.~\ref{fig:z0_mstar_mh}. CGM temperatures as a function of halo mass tend to have a ``U''-shape, with a minimum temperature (relative to $T_{200}$) at halo masses of $\sim\!10^{11-11.5}\,$M$_\odot$. Temperatures in \ark\ runs tend to lie above those of TNG at $z=0$.}
    \label{fig:z0_mstar_Tcgm}
\end{figure}

\begin{figure}
    \centering
    \includegraphics[width=0.99\linewidth]{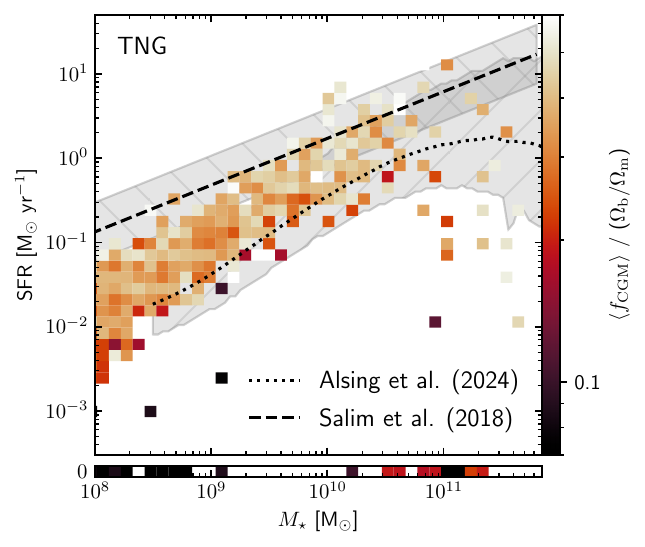}
    \includegraphics[width=0.99\linewidth]{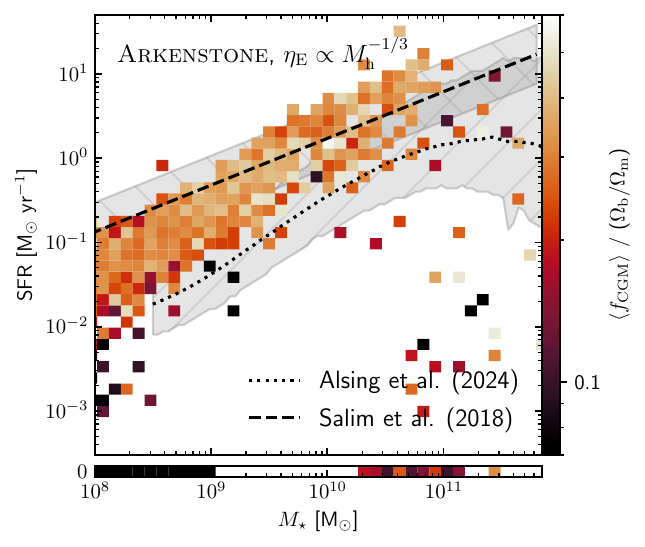}
    \caption{Galaxy SFR as a function of stellar mass at $z=0$ for TNG (top) and the intermediate variable energy loading \ark\ run ($\eta_{\rm E} \propto M_{\rm h}^{-1/3}$, bottom), coloured by the mass-weighted average gas fraction in the CGM (between twice the stellar half mass radius and $R_{200}$), normalised by the universal baryon fraction. The separated axes at the bottom of each panel show the galaxies with a zero SFR. Observational comparison data is from two sources: \citet{Salim2018}, where the median is shown by a dashed black line and the shaded 1-$\sigma$ region has a backwards hash, and \citet{Alsing2024}, which has a dotted median line and a forward hashed 1-$\sigma$ range. With \ark, galaxies tend to have a slightly higher SFR than TNG, but a more suppressed gas fraction in the CGM -- suggesting more preventative feedback.}
    \label{fig:z0MstarSFR}
\end{figure}

In Fig.~\ref{fig:z0_mstar_fgas} we show the median gas fraction within $R_{200}$, relative to the cosmic baryon fraction, as a function of halo mass. The shape of the TNG line is set by the combination of the stellar and BH feedback models. TNG was also calibrated to this quantity, though data exists only for groups (of which there are only a handful in our box) and clusters (of which there are none). As this paper focuses on haloes below this scale, which lies in a different part of parameter space to available observations, we do not show comparison data here. 

Below a halo mass of $\sim\!2\times10^{12}$\,M$_\odot$, the amount of gas in the CGM is primarily regulated by stellar feedback. We have already discussed how TNG tends to rely on ejection of mass from the ISM to regulate star formation. We can see how this ejection also removes much of the CGM at the lowest halo masses, while slightly larger haloes can retain more material. Above halo masses of $\sim\!2\times10^{12}$\,M$_\odot$, the kinetic mode of TNG's AGN feedback typically kicks in, expelling much of the inner CGM. As the halo mass increases further, $R_{200}$ increases and haloes can gravitationally hold onto more of that material, leading to an increase in the gas fraction once more.

A number of differences are present between runs with different energy loadings (top panel), with opposing trends at low and high halo masses. At low masses, the higher energy loadings lead to more preventative feedback, lowering the gas fraction that ends up within $R_{200}$. At high masses, the highest energy loading actually leads to the retention of more gas. This is likely due to the stunted growth of black holes (see Fig.~\ref{fig:z0_mstar_mbh}) leading to less feedback, particularly because fewer BHs will then enter TNG's powerful, BH mass-dependent kinetic mode. 

The mass loading of \ark\ winds makes little difference to the gas fraction of intermediate and high mass haloes for $\eta_\mathrm{M} \le 0.3$. In low mass haloes the mass loading makes more of a difference, once again with lower mass loading leading to more preventative feedback, this time shown by the reduction of the CGM gas fraction for $\eta_\mathrm{M} = 0.1$. The $\eta_\mathrm{M}=0.9$ run has particularly low gas fractions of 10 per cent or below in haloes with masses above $10^{12}$\,M$_\odot$, at least a factor of two below the other runs, though as we show below, this is due to the gaseous halo being converted into (excess) stars.

The winds with variable energy loading smooth the abrupt change in median gas fraction present in TNG at halo masses of $\sim\!10^{10.5}$\,M$_\odot$. Gas fractions within $R_{200}$ at $z=0$ in \ark\ tend to lie below those of TNG -- this CGM depletion can be seen in the regions close to galaxies in Fig.~\ref{Fig:DensRatioMap}. 

In Fig.~\ref{fig:z0_mstar_fbary} we show a similar plot, except now combining both the gas fractions and stellar fractions into the median baryon fraction. Here we see many of the same trends present in the SMHM for haloes below $10^{12}$\,M$_\odot$, with the normalisation and slope of the relation strongly dependent on energy loading, and the mass loading affecting the normalisation. The baryon fraction of the $\eta_\mathrm{M}=0.9$ run at the massive end is very similar to all other runs, despite the significant difference in gas fraction - this run forms far too many stars. The median baryon fractions for the variable energy loading \ark\ runs are very similar to TNG, with only slight differences between the different energy loading slopes.

Preventative feedback can also act through the heating of the CGM (and IGM) around galaxies, inhibiting the cooling of gas onto the central galaxy. To study this, in Fig.~\ref{fig:z0_mstar_Tcgm} we show the mass-weighted temperature of all non-star forming gas cells (those not subject to the eEoS described in Section~\ref{Section:Methods}) within $R_{200}$, normalised by the halo virial temperature $T_{200} = {\mu m_\mathrm{p} G M_{200}}/{(2 k_\mathrm{B} R_{200})}$, where $\mu = 0.59$ is the mean molecular weight of ionised primordial gas. 

The vast majority of haloes in all of the runs in Fig.~\ref{fig:z0_mstar_Tcgm} lie below~1, indicating a sub-virial mean temperature. The CGM temperatures in TNG are closest to the virial temperature at the lowest and highest masses, with a dip in between at $\sim\!10^{11-11.5}\,$M$_\odot$. This dip likely corresponds to haloes with a virial temperature near the peak of the cooling curve, leading to more efficient cooling \citep{Fielding2017}, particularly in TNG where the CGM density tend to be higher than in \ark\ (see Fig.~\ref{fig:z0_mstar_fgas}). Like the gas fraction plot, the relationship at low masses changes more smoothly in \ark\ compared to TNG. The \ark\ runs have a somewhat similar shape, with most runs having a minimum in CGM temperature in the halo mass range $\sim\!10^{11-11.5}\,$M$_\odot$, but with the average $z=0$ temperature being higher than TNG. There are no clear changes between the different \ark\ simulations, with the exception of hotter CGM temperatures for the lowest mass galaxies when the wind's specific energy is increased. We note that this median value of the mass-weighted mean temperatures of haloes hides significant complexity in the CGM, which generally cannot be characterised by a single temperature \citep{Lochhaas2021}. 

In Fig.~\ref{fig:z0MstarSFR} we see how this change in CGM properties drives the regulation of star formation. We show two-dimensional histograms of the SFR of galaxies as a function of stellar mass, highlighting the star-forming main sequence for TNG (top panel) and the variable energy loading \ark\ run with $\eta_\mathrm{E} \propto M_\mathrm{h}^{-1/3}$ (bottom panel). For comparison we also show two observationally inferred main sequences, shown with dotted and dashed black lines \citep{Salim2018,Alsing2024,Sommovigo2025}. The histogram is coloured by the stellar mass-weighted average CGM gas fraction (defined as the gas within $R_{200}$ minus the gas within twice the stellar half mass radius), normalised by the cosmic baryon fraction. The darker regions therefore indicate where the CGM is relatively depleted of gas, which for \ark\ typically occurs in galaxies below the star forming main sequence. TNG tends to have larger CGM gas fractions in general, as also shown in Fig.~\ref{fig:z0_mstar_fgas}. It also tends to have slightly lower SFRs compared to \ark\, though the scatter in the observationally estimated star-forming main sequence precludes a stronger conclusion \citep[see also][]{Donnari2019}.

\subsection{Preventative feedback in action} \label{Section: z2results}

\begin{figure*}
    \centering
    \includegraphics[width=0.8\linewidth]{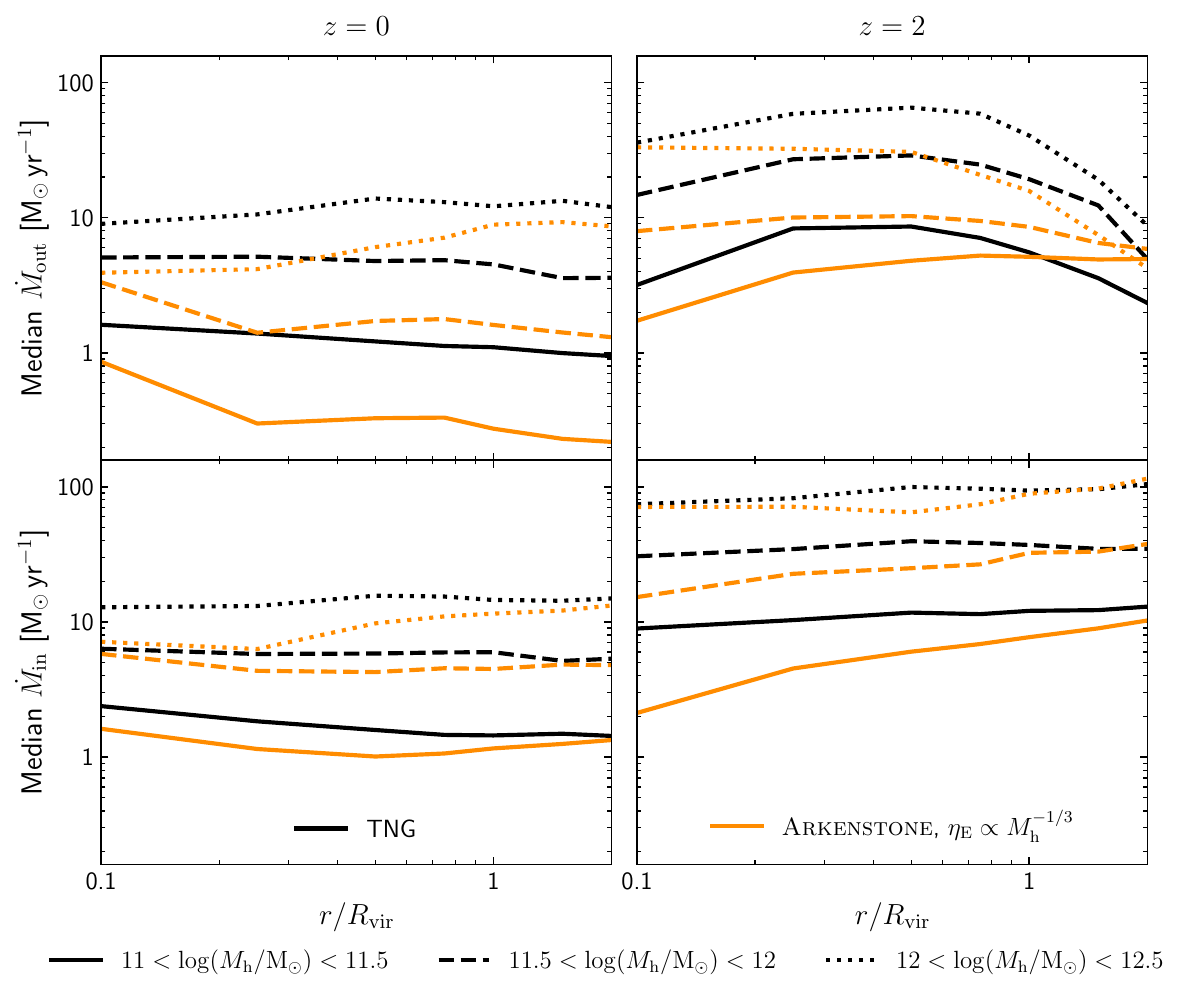}
    \caption{Median mass outflow (top row) and inflow (bottom row) rates as a function of radius, binned in three different halo mass bins. The left and right columns show results at $z=0$ and $z=2$, respectively, with TNG shown in black and the \ark\ variable energy loading run with $\eta_\mathrm{E} \propto M^{-1/3}$ in orange. The mass fluxes in \ark\ are almost always lower than that of TNG, with the gap widening at smaller radii.}
    \label{fig:inflowoutlflow}
\end{figure*}

\begin{figure}
    \centering
    \includegraphics[width=0.8\linewidth]{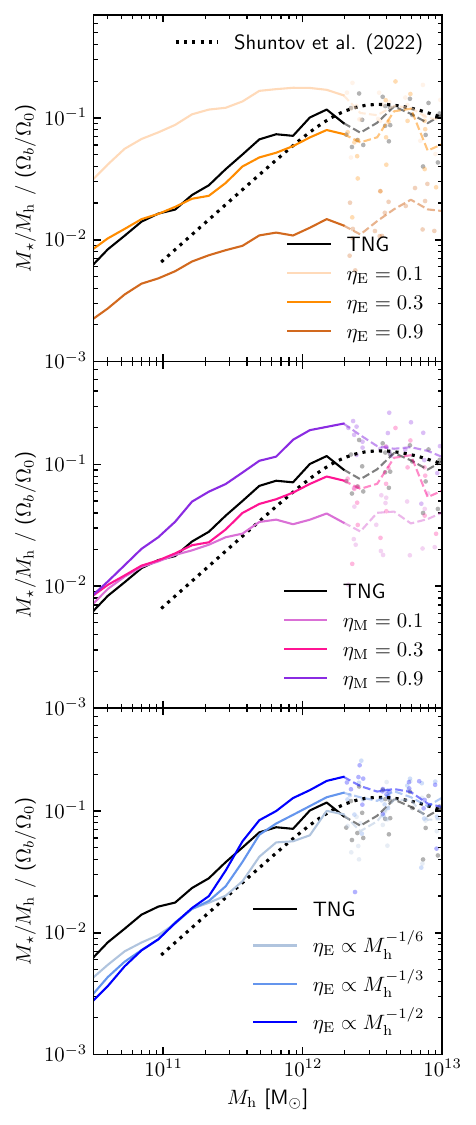}
    \caption{Stellar to halo mass ratios at $z=2$, normalised by the universal baryon fraction \citep[$f_\mathrm{B} = 0.16$,][]{Planck2015Parameters}. Lines and points are displayed in the same way as Fig.~\ref{fig:z0_mstar_mh}, with observational data again from \citet{Shuntov2022}. The mass loadings of the hot winds in \ark\ have an impact on the normalisation of the SMHM relation, with lower values leading to less star formation due to a hotter and faster wind.}
    \label{fig:z2_mstar_mh}
\end{figure}

\begin{figure}
    \centering
    \includegraphics[width=0.8\linewidth]{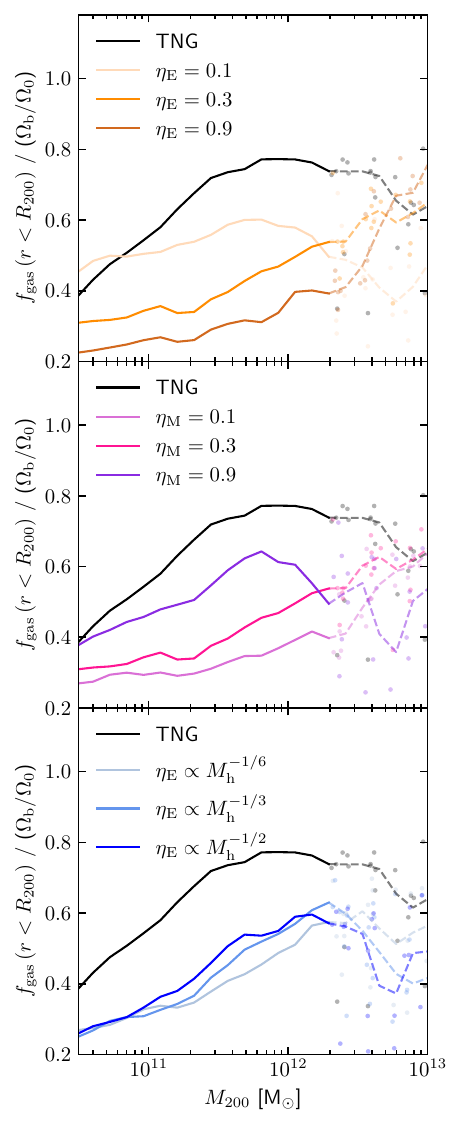}
    \caption{Gas fractions within $R_{200}$ at $z=2$, normalised by the universal baryon fraction \citep[$f_\mathrm{B} = 0.16$,][]{Planck2015Parameters}. Lines and points are displayed in the same way as in Fig.~\ref{fig:z0_mstar_mh}. The amount of gas in the CGM at $z=2$ is significantly different between \ark\ and TNG. The preventative feedback in \ark\ reduces the amount of gas within the virial radius, lowering inflows onto the central galaxy.}
    \label{fig:z2_mstar_fgas}
\end{figure}

\begin{figure}
    \centering
    \includegraphics[width=0.8\linewidth]{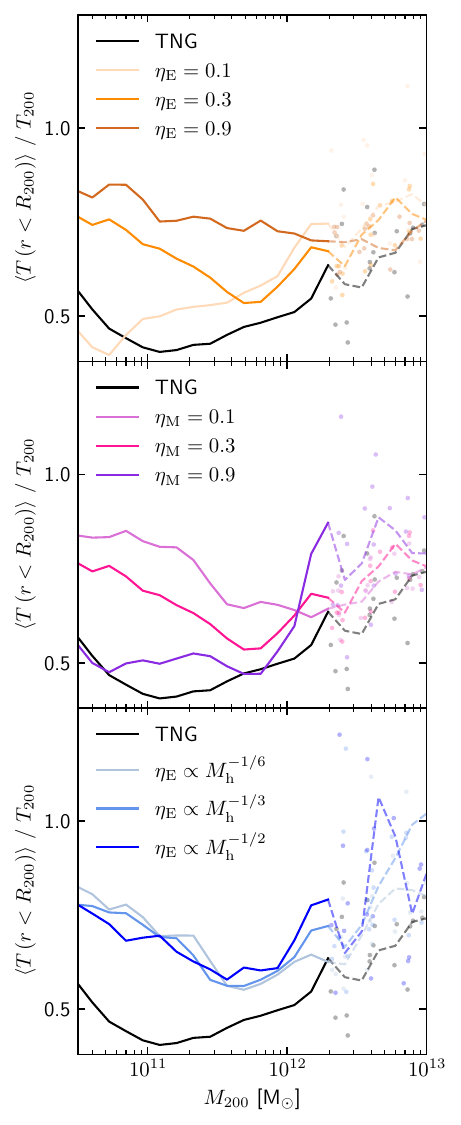}
    \caption{Mass-weighted mean CGM temperatures within $R_{200}$ at $z=2$, normalised by the nominal virial temperature $T_{200}$. Lines and points are displayed in the same way as Fig.~\ref{fig:z0_mstar_mh}. Like the gas fractions, the CGM temperature is very different between \ark\ and TNG. Variations in energy loading significantly change the mean CGM temperature, with a hotter, more diffuse CGM acting to reduce inflows to the central galaxy.}
    \label{fig:z2_mstar_Tcgm}
\end{figure}

In this section we examine the degree to which feedback is ejective vs. preventative in \ark\ simulations, by looking at mass fluxes directly, but also through investigating a subset of the quantities discussed in Sections~\ref{Section: z0results} and \ref{Section: z0results_CGM} at $z=2$. Here, the star formation rate is higher and so the subsequent feedback is stronger. This allows us to illuminate the differences between \textit{how} \ark\ and TNG regulate star formation, even though they both can lead to similar $z=0$ galaxy properties.

Fig.~\ref{fig:inflowoutlflow} shows the total radial outflowing (top row) and inflowing (bottom row) gas mass flow rates as a function of radius, at $z=0$ (left column) and $z=2$ (right column). We calculate the mass fluxes at each radius $r$ by identifying the closest cell to each of $N_\mathrm{pix} = 786,432$ \textsc{healp}ix pixels \citep{HEALPix} and assigning its properties to each pixel. The total mass flow rates then follow as
\begin{equation}
    \dot{M} = \frac{4 \pi r^2}{N_\mathrm{pix}} \sum \rho v_\mathrm{r},
\end{equation}
which we split into outflow (where $v_\mathrm{r}>0$) and inflow rates (where $v_\mathrm{r}<0$). We do this for every central galaxy in each of three mass bins, and present the median value.

Turning first to the outflow rates, we can see how \ark\ has consistently lower mass outflow rates than TNG at both $z=0$ and $z=2$, across all halo masses and radii. The difference between the two is notably smaller than the difference in input loadings (see Fig.~\ref{fig:loadings}) -- the \ark\ winds sweep up a significant mass of material as they move into and through the CGM.

The inflow rates are particularly interesting in the context of preventative feedback. First at $z=2$, where the global SFR is higher and feedback more active, we see that the median mass inflow rates in \ark\ are consistently below those of TNG. The \ark\ runs have the characteristic signature of preventative feedback -- lower inflowing mass fluxes at smaller radii -- showing how gas is prevented from moving towards the galaxy. This is particularly pronounced in the lower mass bin shown in Fig.~\ref{fig:inflowoutlflow}, with a decrease of a factor $4-5$ between $0.1R_\mathrm{200}$ and $2R_\mathrm{200}$. In contrast, the mass inflow in TNG remains largely flat as a function of radius, such that the inflowing mass flux is about 4 times higher at $0.1R_\mathrm{200}$ compared to \ark. A similar trend is seen in the EAGLE simulations, which also has consistently lower inflow rates than TNG \citep{Wright2024}.

At $z=0$ this signature is less clear. We note that the magnitude of the inflowing (and outflowing) fluxes drops between $z=2$ and $z=0$, particularly due to both a lower average SFR and the ejective effect of AGN feedback in TNG's BH model. Nevertheless, \ark\ mass fluxes remain below TNG in all mass bins and at all radii, but often with a narrow gap than at $z=2$. In this \ark\ run, we show in Fig.~\ref{fig:z0MstarSFR} that $z=0$ SFRs lie a little higher than TNG, which could potentially be linked to the narrower difference between \ark\ and TNG in Fig.~\ref{fig:inflowoutlflow} at $z=0$.

In Fig.~\ref{fig:z2_mstar_mh} we show the SMHM at $z=2$, which shows mostly similar trends to $z=0$. Changing the energy loading of the hot wind has the greatest effect on the normalisation of the SMHM (top panel). The mass loading affects the normalisation and the slope of the SMHM at intermediate halo masses (central panel), with higher mass loadings leading to higher stellar masses. As discussed earlier, this relationship is due to lower mass loading causing a hotter and faster wind (with the same total power) and thus increased preventative feedback. The variable energy loadings (bottom panel) counter the overproduction of stars at the low mass end, bringing the SMHM closer to the observational estimates. Notably, most of the simulations lie above the observationally inferred SMHM at $z=2$ from \citet{Shuntov2022}, though the \ark\ variable energy loading runs are close.

The most significant differences between the runs, which hold the key to understanding how \ark\ differs from TNG, are shown in the CGM gas fraction (Fig.~\ref{fig:z2_mstar_fgas}) and temperature (Fig.~\ref{fig:z2_mstar_Tcgm}) plots. 

Turning first to the gas fractions, for TNG this starts off low at the smallest halo masses, before increasing to $\sim\!80$ per cent of the cosmic baryon fraction at a halo mass of $10^{12}$\,M$_\odot$, and slowly declining beyond that as the black hole feedback takes effect. For \ark\, the gas fraction increase has a much shallower slope across halo masses. For simulations with different constant energy loading values (top panel), the normalisation for haloes below $10^{12}$\,M$_\odot$ is inversely proportional to $\eta_\mathrm{E}$. At the TNG `turnover' mass of $10^{12}$\,M$_\odot$, the \ark\ runs have a gas fraction of 30-50 per cent of baryons, around half of TNG's value at this point. The strong implication of this is a reduction in halo-scale accretion rates - either due to high specific energy winds pushing gas outside $R_{200}$, or the prevention of gas accreting within $R_{200}$ in the first place. This `prevention-via-CGM-depletion' scenario, in contrast to ISM ejection, is a central feature of the \ark\ model. Notably, the relationship between energy loading and gas fraction reverses in haloes above $10^{12}$\,M$_\odot$, though this is likely due to the variations in stellar feedback affecting black hole growth and feedback.

Varying the mass loading also affects the gas fractions (central panel) - the lowest mass loading of $\eta_\mathrm{M} = 0.1$ also has the lowest gas fractions for low and intermediate mass galaxies, indicating that hotter and faster winds reduce the baryon content of the CGM even more at higher redshift. The variable energy loadings on the other hand lead to a gradually increasing gas fraction as a function of halo mass, up to a halo mass of $\sim\!10^{12}$\,M$_\odot$. This slope is shallower than in TNG, and the median gas fraction values are below TNG at all halo masses.

The temperature of the CGM is also an indicator of preventative feedback, with hotter, diffuse gas having longer cooling times and so reducing inflows onto the central galaxy. The CGM temperature at $z=2$ is shown in Fig.~\ref{fig:z2_mstar_Tcgm}, defined in the same way as Fig.~\ref{fig:z0_mstar_Tcgm}. Here we again see a strong difference between TNG and \ark. The CGM temperatures for haloes around $\sim\!10^{11}\,$M$_\odot$ are much cooler in TNG than in most \ark\ runs, and the \ark\ runs stay hotter throughout the entire halo mass range.

The temperature varies with both energy and mass loading - higher \textit{specific} energy winds lead to a hotter and more diffuse CGM, more able to prevent gas accreting onto galaxies and forming stars. The highest mass loaded \ark\ run leads to a cooler CGM closer to TNG, as more energy is used to transport a heavier wind. The slope of the variable energy loading does not strongly impact the CGM temperature at $z=2$.

Across the parameter space studied in this work, the CGM of galaxies in \ark\ is hotter and more diffuse, an indicator of the large role preventative feedback plays in regulating star formation. 

\subsection{Evolution with redshift} \label{Section:RedshiftEvo}

\begin{figure}
    \centering
    \includegraphics[width=0.8\linewidth]{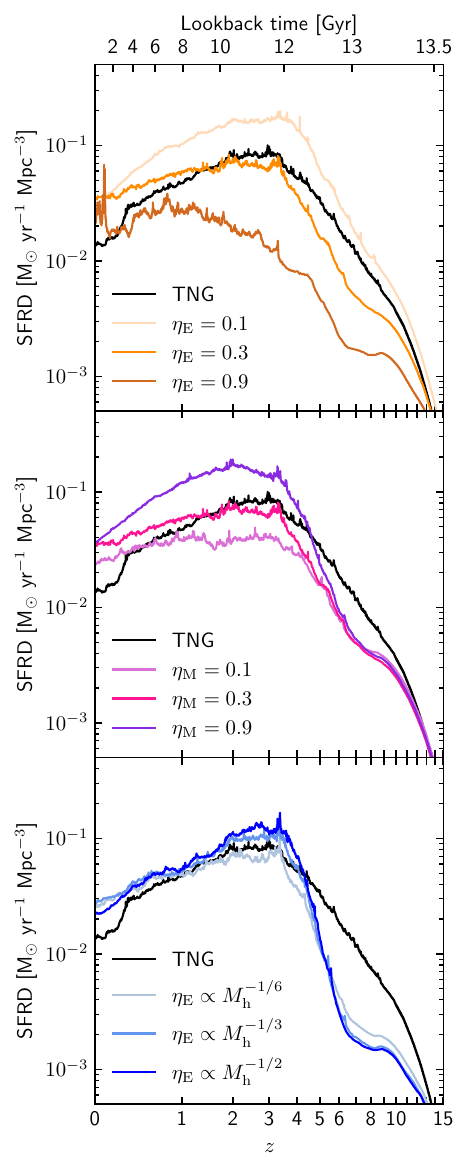}
    \caption{Global star formation rate density at each simulation timestep in our simulated boxes as a function of redshift. Line styles are the same as those in Fig.~\ref{fig:z0_mstar_mh}. It is not trivial to make observational comparisons to this quantity, so we postpone this to future work. Higher energy loadings can lead to much more suppression in star formation.}
    \label{fig:SFRD}
\end{figure}

To emphasise the differences in redshift evolution between the runs we plot the global star formation rate density (SFRD) versus redshift in Fig.~\ref{fig:SFRD}. It should be noted that a comparison to observations is not trivial to make, not least because of the limited box size, so we omit this here. However, the relative comparison between simulations remains informative. In these panels, it is clear that cosmic star formation history is most sensitive to the change in energy loading. 

In the top panel, with fixed mass loadings of $\eta_{\rm M} =0.3$, the run with the lowest energy loading, $\eta_\mathrm{E}=0.1$ has a global SFRD that rises higher than any other run at cosmic noon, and its turnover occurs slightly earlier than in TNG, though we caution this is dominated by the largest few haloes in the box. These are in turn sensitive to the interaction between stellar feedback and the TNG BH model, which again has not been re-calibrated and is beyond the scope of this work. The highest energy loading has a significantly lower SFR throughout cosmic history - especially at high redshift and down to cosmic noon.

For runs with a fixed energy loading $\eta_{\rm E} = 0.3$, varying the mass loading of the hot winds also affects the SFRD, though for high and intermediate redshifts the effect is smaller than that of energy loading. Again, lower mass loadings suppress star formation further, due to the wind being hotter and faster \citep[e.g][]{Voit2024a}.

The SFRDs for runs with variable energy loading have a different shape to those from TNG, with a suppression of star formation at high redshift, before a rapid increase to cosmic noon. This is due to the higher energy loading for lower halo masses, which suppress star formation at early times. As discussed in Section \ref{Section:Discussion}, the tests we have performed in this work have not been calibrated, and are an investigation of how high specific energy winds \textit{alone} can regulate star formation via preventative feedback. To fully explain the number densities of the UV luminous early galaxies identified with \textit{JWST} \citep[e.g.][]{Bunker2023,Curtis-Lake2023,Tacchella2023,Finkelstein2024} may require a change in the star formation prescription and possibly a combination of ejective and preventative feedback, which we will study in detail in future work with the full \ark\ model.

\section{Discussion} \label{Section:Discussion}

The core result of this paper is that star formation can be regulated by high specific energy winds, with low input mass loadings and sub-unity energy loadings. Fig.~\ref{fig:z0_SMF} shows this across a wide variety of halo masses using such winds \textit{only} (as a reminder, the cold cloud component of the full \ark\ framework is not active in this paper). This is a significant pivot away from the highly mass loaded winds of many existing cosmological simulations -- a shift from primarily ejective towards primarily preventative feedback. The balance between the two has yet to be determined, and will require cosmological simulations with the full \ark\ model, but we have shown in this work how high mass loadings are \textit{not necessary} to regulate star formation in low and intermediate mass galaxies if sufficient energy is injected into the system via hot winds (controlled by the ratio of energy loading to mass loading). Further, less easily accessibly observables, especially of the CGM, will be essential to disentangle the contribution from ejective and preventative feedback in the future.

\ark\ has low mass loadings, but importantly also does not require unphysically large amounts of energy to be injected. Indeed we emphasise that \ark\ is green: star formation is regulated while using considerably less energy than existing simulations. As discussed in Section \ref{Section:Loadings}, at $z=0$ the mass loadings of galaxies at $10^{10}$, $10^{11}$, and $10^{12}$\,M$_\odot$ in the TNG model are approximately 31.4, 10.2 and 1.5, respectively. For energy loadings the values in TNG in haloes of $10^{10}$, $10^{11}$, and $10^{12}$\,M$_\odot$ are approximately 3.6, 1.36 and 0.93, respectively. Compare these to the fixed $\eta_\mathrm{M} = \eta_\mathrm{E} = 0.3$ of the central fixed variation of \ark\, or even the variable energy loading run which has a cap at $\eta_\mathrm{E,max} = 1$, and we can see a dramatic shift in how galaxies are regulated -- much less material enters and leaves the galaxy with \ark's preventative feedback. We also note that EAGLE requires an enhanced energy injection (likely somewhat higher than what is nominally available from supernovae) to regulate low mass galaxies \citep[see e.g.][]{EAGLE2}. As discussed in Section~\ref{Section:Loadings}, the actual amount of energy available to drive winds out of the galaxy will be smaller than $\eta_\mathrm{E,max}$, as energy will be needed to escape the ISM. Energy loadings larger than unity are likely to be unphysical, but the true amount of energy available to drive winds must be investigated in small-scale simulations.

Looking further at the gas fractions at $z=2$, it is clear that the CGM is significantly depleted within $R_{200}$ in the \ark\ runs, considerably more so than in TNG. Interestingly, however, by $z=0$ the differences between the two are much smaller -- much of the ejected material is recaptured by the halo. Despite this, at $z=0$ there is still a correlation between CGM gas fraction and SFR, as shown in Fig.~\ref{fig:z0MstarSFR}. These results have numerous implications for metal enrichment of the CGM and IGM, and for gas flows since cosmic noon, which will be studied in future work. 

In Sections~\ref{Section:Intro} and \ref{Section:Methods} we motivated our investigation and choices of parameters based on recent analytic and semi-analytic works. The gas-regulator model of \citet{Carr2023} simplifies the baryon cycle to an exchange of mass, metals, and energy between reservoirs of stars, the ISM and the CGM. They find that observables such as the SMHM relation can be reproduced with winds that have low mass loadings and high specific energy. They find that $z=0$ stellar masses are largely insensitive to changes in mass (and metal) loading, but dependent on energy loading, in good agreement with our results. While the dependence of $\eta_{\rm E}$ on halo mass that \citet{Carr2023} find is needed to reproduce the SMHM is slightly steeper than what we have tested, we find the same general trend of higher $\eta_{\rm E}$ at lower halo masses is required. A notable result of \citet{Carr2023} is that highly mass loaded winds tend to lead to a denser CGM with lower specific energy, allowing gas to return to the galaxy on shorter timescales. Our results on the CGM gas fraction and temperature support this, with the highly mass loaded winds of TNG generally leading to a cooler, more massive CGM (see Figs.~\ref{fig:z0_mstar_fgas}, \ref{fig:z0_mstar_Tcgm}, \ref{fig:z2_mstar_fgas}, and \ref{fig:z2_mstar_Tcgm}). 

\citet{Pandya2023} used a semi-analytic model framework to study CGM evolution and its impact on galaxy evolution, with the additional inclusion of turbulence in addition to thermal energy. Consistent with the work presented here, \citet{Pandya2023} find that the average global CGM temperature does not have to be at the virial temperature (and is in fact often below). They also find higher specific energy winds (with higher $\eta_{\rm E}/\eta_{\rm M}$) to lead to a hotter, more turbulent CGM. We do find a similar temperature increase at $z=2$ (Fig.~\ref{fig:z2_mstar_Tcgm}), though the trend is less clear at $z=0$. Interestingly, with \ark\ the ratio $\eta_{\rm E}/\eta_{\rm M}$ alone does not set the average CGM temperature, as the runs with ($\eta_{\rm E}$, $\eta_{\rm M}$) = (0.9, 0.3) and (0.3, 0.1) tend to have slightly different temperatures (with the former being hotter) despite the same loading ratio. We postpone a detailed study of the role of CGM turbulence to future work. 

A key comparison is to the minimalist regulator model of \citet{Voit2024a,Voit2024b}, which reduces galaxy evolution and the baryon cycle to three coupled differential equations. 
This model predicts that the CGM expands and contracts depending on the ratio between the specific energy of gas injected into the CGM via feedback versus the specific energy it had when accreted. Thus, the minimalist regulator model offers a physical interpretation for many of the results in this paper -- high specific energy winds drive the expansion of the CGM, reducing the inflow rate onto central galaxies and the corresponding SFR. This is particularly evident in Fig.~\ref{Fig:DensRatioMap}, which highlights the expanded size of the CGM around many galaxies in our simulated volume. The regulator model also offers an explanation for why our \ark\ simulation with the highest mass loading fails to regulate star formation - a higher $\eta_{\rm M}$ at fixed $\eta_{\rm E}$ reduces the specific energy of the wind, leading to a contracted CGM, more cooling, more recycling of material, and thus a higher SFR. Our simulations provide evidence for the important role of the specific energy of galactic winds in regulating star formation and shaping the properties of the CGM.

\section{Conclusions} \label{Section:Conclusions}

In this paper we have demonstrated the use of the new \ark\ galactic wind model in cosmological simulations for the first time. We described the refinement scheme required for the \textsc{Arkenstone-Hot} model to work in cosmological simulations, before testing it on a ($36.9\,$Mpc)$^3$ box. We perform a series of numerical experiments, with both fixed mass and energy loadings and variable energy loadings that scale with DM velocity dispersion (a proxy for halo mass). In this work we do not calibrate our results to observations, but simply explore the impact of changing the loadings. Throughout the paper, we present our results in comparison with simulations run with the TNG model for reference.

Our key conclusions are as follows:
\begin{enumerate}

\item The \ark\ simulations presented in this paper have substantially lower mass and energy loadings in galactic winds compared to the existing TNG model. This increases the specific energy of galactic winds -- they become hotter, faster, and lighter.

\item Such winds do not remove large amounts of gas from the ISM of galaxies, but instead regulate star formation by preventing accretion onto galaxies in the first place via the heating, expansion, and depletion of the CGM, particularly at high redshift.

\item The energy loading is the key parameter which affects the stellar masses and sizes of galaxies and properties of the CGM, with a much smaller change arising from variations in mass loading -- both the total energy injected into the CGM and the specific energy of the wind contribute to these results.

\item With a variable energy loading that increases for lower mass galaxies, \ark\ provides reasonable agreement with observables like the stellar mass-halo mass relation, and does so with a substantially different feedback paradigm to TNG.

\item \ark\ can match observed galaxy scaling relations without exceeding the supernovae energy budget and with significantly lower mass loadings than required in other cosmological simulations.

\end{enumerate}

Future work with \ark\ will delve further into the new parameter space opened up by the inclusion of detailed modelling of high specific energy winds. An important aspect of the \textsc{Arkenstone-Hot} model to be refined further is the metal loading of the hot winds (discussed in Appendix \ref{Section:MetalAppendix}), which will have a significant impact on observables of the CGM and IGM. Additionally, while in this work we tied wind loadings to approximately halo--scale properties (enabling a simple numerical experiment), in future we will instead base them on local ISM properties (informed by small scale ISM patch simulations), increasing predictive power. Another key advance to be made with \ark\ is the utilisation of the full multiphase wind model \citep{Smith2024b} in cosmological volumes, with subgrid cool clouds embedded in the hot wind. Only by using both aspects of the model can we begin to determine the balance between ejective and preventative feedback in galaxy evolution.

\section*{Acknowledgements}
The authors thank the anonymous referee for helpful comments that have led to an improved paper. The authors would like to thank Shy Genel for providing the initial conditions used in these simulations, and Chris Carr and Mark Voit for useful conversations. This work was supported by the Simons Collaboration on “Learning the Universe”. The authors wish to thank the synthetic observations working group of the Learning the Universe collaboration for assembling observational comparison data, and Stephen Thorp and Justin Alsing for sharing data from \texttt{pop-cosmos} \citep{Alsing2024}. 
Computations were performed on the HPC systems Iron, at the Flatiron Institute; Orion, at the Max Planck Computing and Data Facility (MPCDF); Anvil, at Purdue University (accessed through allocation PHYS240043 from the Advanced Cyberinfrastructure Coordination Ecosystem: Services \& Support (ACCESS) program, which is supported by National Science Foundation grants \#2138259, \#2138286, \#2138307, \#2137603, and \#2138296); and Frontera (accessed through Leadership Resource Allocation AST20007), at the Texas Advanced Computing Center \citep{Frontera,Anvil,ACCESS}. 
GLB acknowledges support from the NSF (AST-2108470, AST-2307419), NASA TCAN award 80NSSC21K1053, and the Simons Foundation. The Flatiron Institute is supported by the Simons Foundation. 
This work made use of the \textsc{numpy} \citep{Numpy}, \textsc{scipy} \citep{SciPy}, \textsc{matplotlib} \citep{Matplotlib}, and \textsc{cmasher} \citep{CMasher} \textsc{python} packages.

\section*{Data Availability}
The data used in this work will be shared on reasonable request to the corresponding author.







\appendix

\section{Metallicity of galactic winds} \label{Section:MetalAppendix}

\begin{figure}
    \centering
    \includegraphics[width=\linewidth]{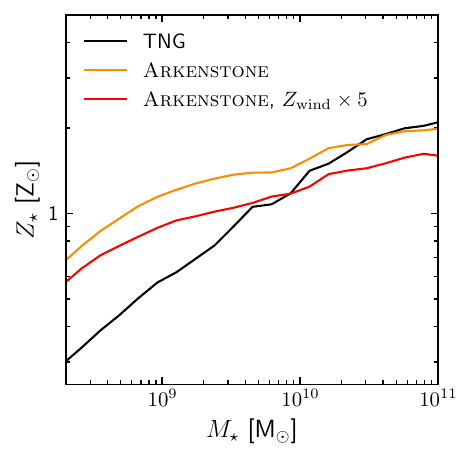}
    \caption{Metallicity of stars within twice the stellar half mass radius, as a function of stellar mass. We show TNG (black) and our central fixed loading \ark\ run ($\eta_\mathrm{E} = \eta_\mathrm{M} = 0.3$ (orange), and additionally show the same \ark\ run but with a the wind metallicity enhanced by a factor of 5 (red).}
    \label{fig:StellarMetals}
\end{figure}

\begin{figure}
    \centering
    \includegraphics[width=\linewidth]{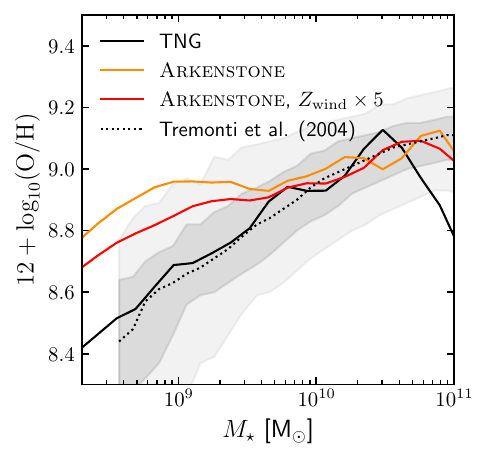}
    \caption{Median gas metallicity (within twice the stellar half mass radius) as a function of stellar mass, with the line styles the same as Fig.~\ref{fig:StellarMetals}. For comparison we show observational data from \citet{Tremonti2004}, with the dotted line showing the median, and dark and light grey shaded regions encompassing the 1- and 2-$\sigma$ scatter, respectively.}
    \label{fig:GasMetals}
\end{figure}

\begin{figure}
    \centering
    \includegraphics[width=\linewidth]{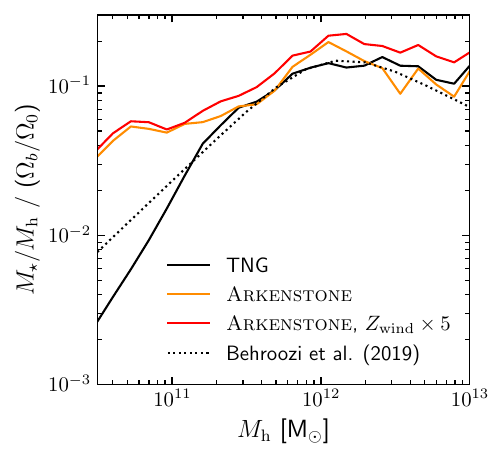}
    \caption{Stellar mass - halo mass relation for TNG, \ark\ with fixed $\eta_\mathrm{E}=\eta_\mathrm{M}=0.3$, and the same \ark\ run with wind metallicity enhanced by a factor of 5. The stellar mass of galaxies does not change significantly with variations in wind metallicity, especially compared with variations due to loading parameters. }
    \label{fig:SMHM_Metals}
\end{figure}

The metal content of galaxies is an important diagnostic for the amount of material produced by stars that remains within the galaxy as it evolves, as metals created in a galaxy can only be locked into future stars or leave via outflows. Stellar metallicity provides an integrated version of this, with enriched gas remaining in a galaxy forming an enriched generation of stars. Gas metallicity is a more variable measurement, and can be affected by both enriched outflows and fresh inflows from the IGM.

The mass loading of winds directly affects the metallicity of a galaxy, as it determines how much gas is removed from the ISM. The very high mass loadings of TNG in the dwarf regime therefore have to be offset by `wind stripping', whereby metals are removed from wind particles at launch and distributed over neighbouring ISM cells, to decrease the metallicity of the wind and avoid over-depleting the ISM. The fraction of metals stripped as wind particles leave the galaxy in TNG is a parameter of the model, and provides a reasonable match to the observed mass-metallicity relation. This is shown in black for stellar metallicity in Fig.~\ref{fig:StellarMetals}. Gas metallicity is displayed in Fig.~\ref{fig:GasMetals}, calculated in an analogous way to observations, as the ratio of oxygen to hydrogen abundances in star-forming gas cells in the simulation \citep[without assuming oxygen makes up 35 per cent of metal mass, unlike][]{Torrey2019}.

In \ark\ we have the opposite problem, where the low (and fixed) mass loadings tested in this paper do not remove enough metals from the ISM to match the observed mass-metallicity relation, shown in orange in Figs.~\ref{fig:StellarMetals} and \ref{fig:GasMetals}. Both figures show that our fixed loading \ark\ run is close to TNG at the massive end, but significantly overestimates metal masses in the dwarf regime. We note that the \ark\ metallicity plots are will already be higher than TNG due to a larger number of stars producing more metals.

This is not unexpected, as the current implementation ejects mass with the same metallicity as the ISM; however, one might expect the hot outflows produced by supernovae to be much more enriched. Indeed, high-resolution simulations of small ISM regions find that a significant fraction of the highly enriched SN ejecta is directly launched into the wind \citep[e.g.,][]{Hu2019,Kim2020a,Steinwandel2024b}, such that the metallicity of the hot outflow is closer to constant rather than being directly proportional to the ISM metallicity \citep[e.g.,][]{LiBryan2020,Kim2020b}. Although we do not explore such a model in this work, we do demonstrate that the predicted mass-metallicity relation is sensitive to this assumption. In particular, we also show (in these figures) a run where we enhance the metallicity of the \ark\ wind by a factor of 5.
This lowers the normalisation of the stellar metallicities in Fig.~\ref{fig:StellarMetals}, but does not notably affect the slope. The gas metallicities in Fig.~\ref{fig:GasMetals} are lowered in low-mass galaxies while remaining largely unchanged (and consistent with observations) at the massive end. This change does not significantly affect any other galaxy properties, seen in the SMHM in Fig.~\ref{fig:SMHM_Metals}, suggesting our other results are robust to changes in the metallicity of the wind. Note that a (nearly) constant wind metallicity would correspond to a wind metallicity enhancement factor (relative to the ISM metallicity) which increases with decreasing stellar mass, a trend that would bring our predictions into closer agreement with observations.  We postpone a full study of the metal content of galaxies and the impact of varying mass loading and metal enhancements to future work, where we will simultaneously investigate the metal content of the ISM, CGM and IGM.



\clearpage
\twocolumn[{%
\vspace*{17pt}
\begin{center}
{\Large\bfseries Correction to: Prevention is better than cure? Feedback from high specific energy winds in cosmological simulations with \textsc{Arkenstone}}

\vspace{14pt}
{\large Jake S. Bennett\orcidlink{0000-0002-8573-2993},$^{1}$
Matthew C. Smith\orcidlink{0000-0002-9849-877X},$^{2}$
Drummond B. Fielding\orcidlink{0000-0003-3806-8548},$^{3,5}$
Greg L. Bryan\orcidlink{0000-0003-2630-9228},$^{4,5}$
\\
Chang-Goo Kim\orcidlink{0000-0003-2896-3725},$^{6}$
Volker Springel\orcidlink{0000-0001-5976-4599},$^2$ 
Lars Hernquist,$^1$
Rachel S. Somerville\orcidlink{0000-0002-6748-6821},$^5$
and Laura Sommovigo\orcidlink{0000-0002-2906-2200}$^5$}

\vspace{10pt}
{\footnotesize\begin{minipage}{0.95\textwidth}\centering
$^{1}$Center for Astrophysics | Harvard \& Smithsonian, 60 Garden Street, Cambridge, MA 02138, USA\\%
$^{2}$Max-Planck-Institut f{\"u}r Astrophysik, Karl-Schwarzschild-Str. 1, D-85748, Garching, Germany\\%
$^{3}$Department of Astronomy, Cornell University, Ithaca, NY 14853, USA\\%
$^{4}$Department of Astronomy, Columbia University, 550 West 120\textsuperscript{th} Street, New York, NY 10027, USA\\%
$^{5}$Center for Computational Astrophysics, Flatiron Institute, 162 5\textsuperscript{th} Avenue, New York, NY 10010, USA\\%
$^{6}$Department of Astrophysical Sciences, Princeton University, 4 Ivy Lane, Princeton, NJ 08544, USA\\%
\end{minipage}}
\end{center}
\vspace{14pt}
}]
\thispagestyle{plain}
\setcounter{figure}{0}
\setcounter{table}{0}
\renewcommand{\thefigure}{E\arabic{figure}}
\renewcommand{\thetable}{E\arabic{table}}

\begin{keywords}
errata, addenda -- methods: numerical -- galaxies: formation -- galaxies: evolution -- hydrodynamics
\end{keywords}




This is a correction to our paper \textit{Prevention is better than cure? Feedback from high specific energy winds in cosmological simulations with \textsc{Arkenstone}}, which was recently published in MNRAS Volume 543, Issue 2, pp. 1456-1478 (2025). In brief, we correct a coding error in the thermal energy injection during wind-particle recoupling, which introduced an unintentional redshift dependence in the energy loading. With this fixed, we recover similar $z=0$ galaxy properties by increasing the global energy loading by a factor of $\sim\! 2-3$ and removing the cap at $\eta_\mathrm{E}=1$. The main conclusions of the paper remain unchanged, although some high-redshift results (in particular fig.~13) should now be interpreted with caution. We detail the error, introduce new simulations with the error fixed, and present the corrected figures from these new simulations below.

\hspace{3cm}

After publication, an error was discovered in the simulation code for the recoupling of wind particles in the circumgalactic medium (CGM), which meant that, at scale factor $a$, we were accidentally injecting a factor of $a^{-2}$ too much thermal energy each time a full displacement recoupling occurred (the kinetic energy input was unaffected). While the energy loadings shown in the original Fig.~1 were correct at $z=0$, they hid a significantly higher energy loading at higher redshift. 

With this error corrected, we ran test simulations and found that we can achieve similar $z=0$ results to our original work, though this requires overall energy loadings approximately a factor of two higher than those stated in the original paper once the accidental redshift dependence was removed. Given the computational expense involved in running the full suite we have not re-run every simulation with the corrected code, but we present below results from a variable energy loading run (with $\eta_\mathrm{E} = 0.9[\sigma_{\mathrm{DM,1D}} / 51\,\mathrm{km\,s}^{-1}]^{-3/2}\approx 0.9 [M_\mathrm{h} / 10^{11}\,\mathrm{M}_\odot]^{-1/2}$). 

\begin{figure}
    \centering
    \includegraphics[width=0.8\linewidth]{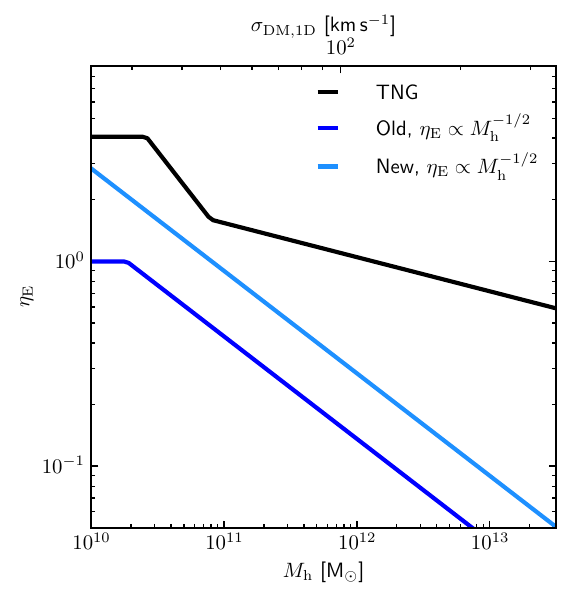}
    \caption{Corrected bottom panel of fig.~1 of \citet{err:Bennett2025}, showing input wind energy loading as a function of halo mass and dark matter velocity dispersion. The TNG and old \textsc{Arkenstone} lines show values at $z=0$, but both have higher loadings at higher redshift. The new \textsc{Arkenstone} simulation has this fixed scaling at all redshifts.}
    \label{err:Fig:EnergyLoading}
\end{figure}

In Fig.~\ref{err:Fig:EnergyLoading} we show the variable energy loading (with a slope of approximately $\eta_\mathrm{E} \propto M_\mathrm{h}^{-1/2}$) of our new run in light blue (middle line), compared to TNG in black (upper line). For reference we also show the line for the equivalent run as shown in the original fig.~1 (bottom line), though we note this line (as well as the line for TNG) is only true for $z=0$.

\begin{figure}
    \centering
    \includegraphics[width=0.8\linewidth]{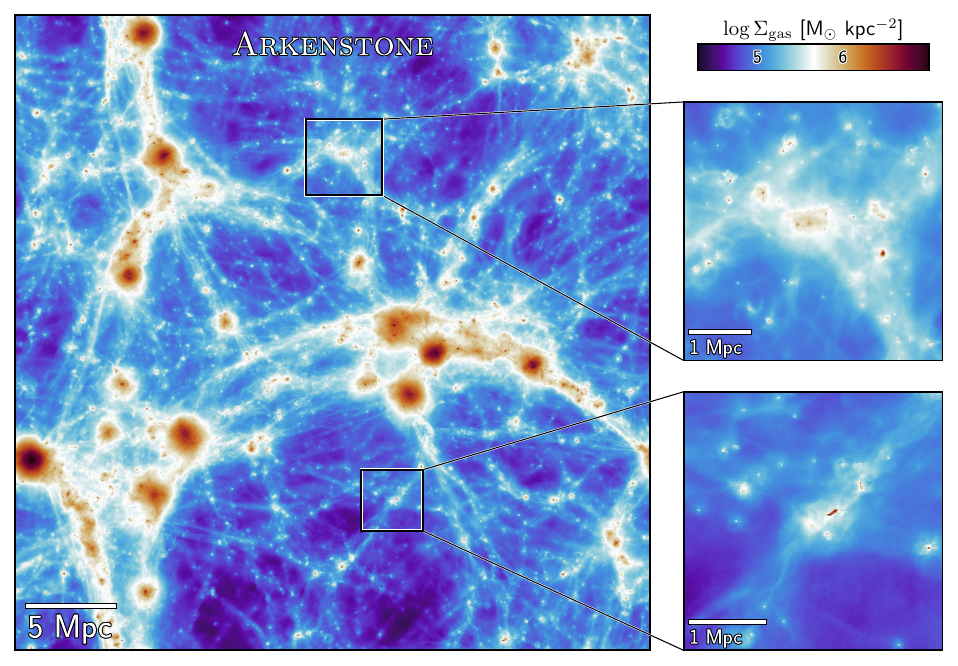}
    \caption{Corrected top panel of fig.~2 of \citet{err:Bennett2025}, showing the gas surface density map of the entire simulation.}
    \label{err:Fig:DensMap}
\end{figure}

\begin{figure*}
    \centering
    \includegraphics[width=0.8\linewidth]{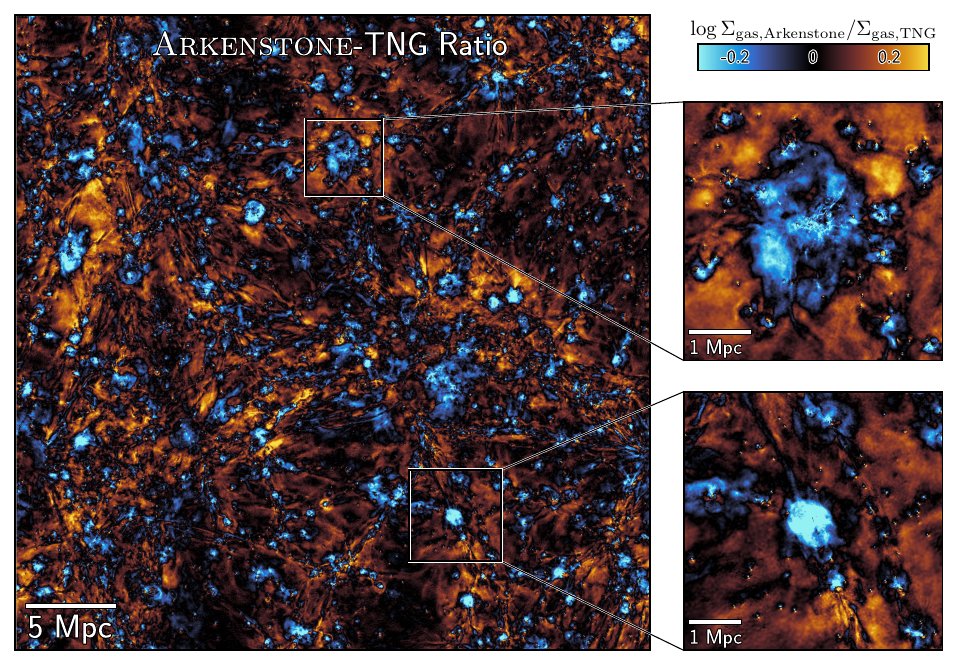}
    \caption{Corrected fig.~3 of \citet{err:Bennett2025}, showing the ratio of the surface density map shown in Fig.~\ref{err:Fig:DensMap} with the equivalent map with the TNG model.}
    \label{err:Fig:DensRatioMap}
\end{figure*}

\begin{figure}
    \centering
    \includegraphics[width=0.8\linewidth]{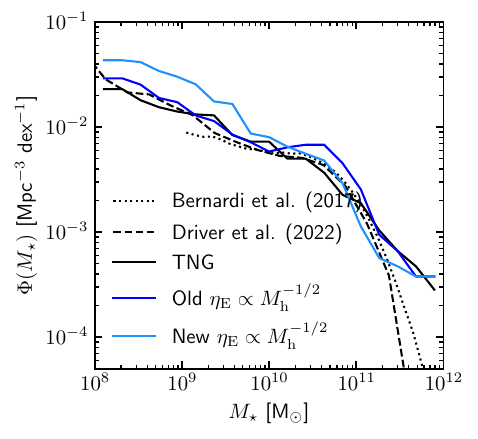}
    \caption{Corrected bottom panel of fig.~4 of \citet{err:Bennett2025}, showing the stellar mass function at $z=0$.}
    \label{err:Fig:z0SMF}
\end{figure}

\begin{figure}
    \centering
    \includegraphics[width=0.8\linewidth]{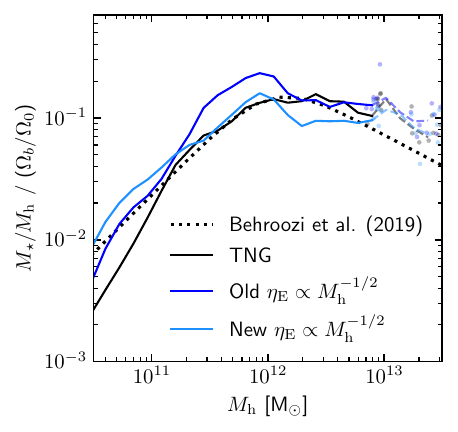}
    \caption{Corrected bottom panel of fig.~5 of \citet{err:Bennett2025}, showing the median stellar mass to halo mass relation normalised by the cosmic baryon fraction.}
    \label{err:Fig:SMHM}
\end{figure}

\begin{figure}
    \centering
    \includegraphics[width=0.8\linewidth]{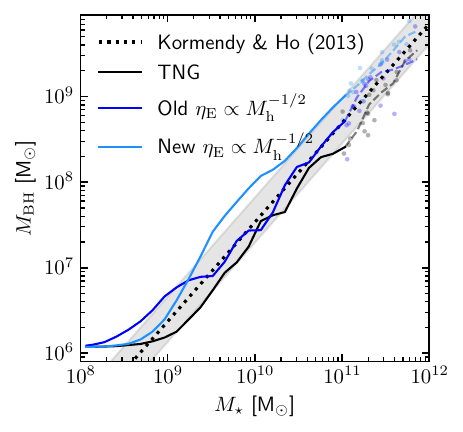}
    \caption{Corrected bottom panel of fig.~6 of \citet{err:Bennett2025}, showing the black hole - stellar mass relation.}
    \label{err:Fig:mstar_mbh}
\end{figure}

\begin{figure}
    \centering
    \includegraphics[width=0.8\linewidth]{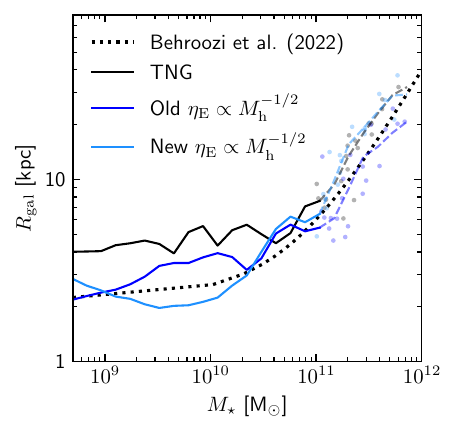}
    \caption{Corrected bottom panel of fig.~7 of \citet{err:Bennett2025}, showing the median stellar half mass radius as a function of stellar mass.}
    \label{err:Fig:Rgal}
\end{figure}

\begin{figure}
    \centering
    \includegraphics[width=0.8\linewidth]{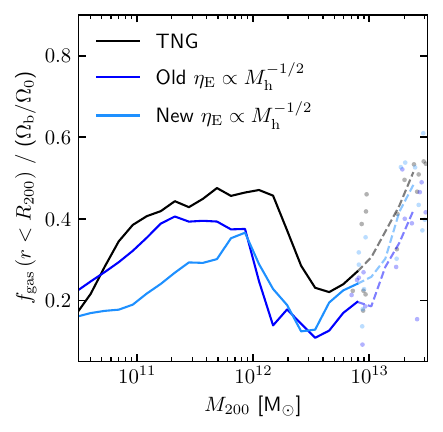}
    \caption{Corrected bottom panel of fig.~8 of \citet{err:Bennett2025}, showing the median gas fractions within $R_{200}$ at $z=0$ normalised by the cosmic baryon fraction.}
    \label{err:Fig:z0_mstar_fgas}
\end{figure}

\begin{figure}
    \centering
    \includegraphics[width=0.8\linewidth]{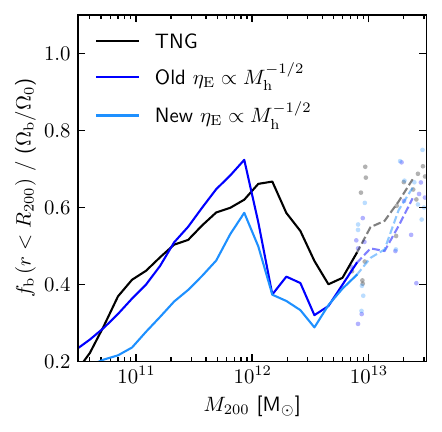}
    \caption{Corrected bottom panel of fig.~9 of \citet{err:Bennett2025}, showing the median baryon fractions within $R_{200}$ at $z=0$ normalised by the cosmic baryon fraction.}
    \label{err:Fig:z0_mstar_fbary}
\end{figure}

\begin{figure}
    \centering
    \includegraphics[width=0.8\linewidth]{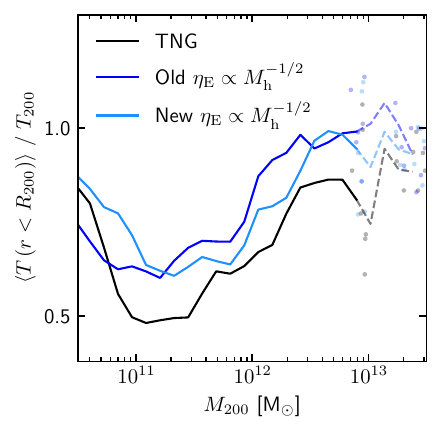}
    \caption{Corrected bottom panel of fig.~10 of \citet{err:Bennett2025}, showing the mass-weighted mean CGM temperature within $R_{200}$ at $z=0$ normalised by the virial temperature.}
    \label{err:Fig:z0_mstar_TCGM}
\end{figure}

\begin{figure}
    \centering
    \includegraphics[width=\linewidth]{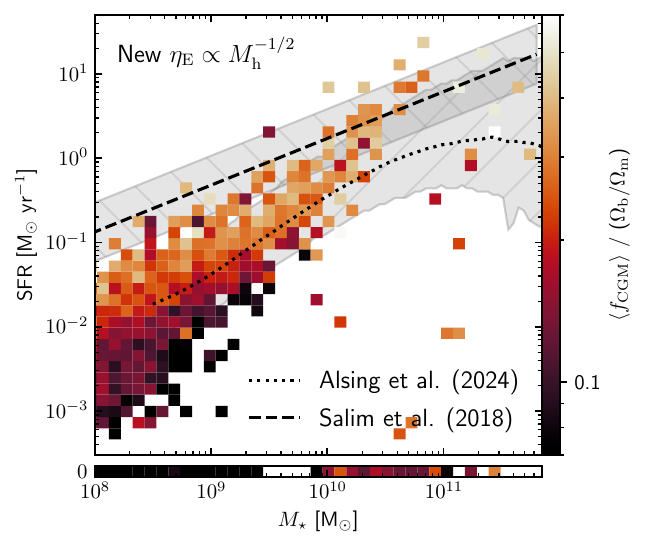}
    \caption{Corrected bottom panel of fig.~11 of \citet{err:Bennett2025}, showing galaxy SFR as a function of stellar mass at $z=0$, coloured by the mass-weighted average CGM gas fraction. We also show the star forming main sequence from observations \citep[with data from][]{err:Salim2018,err:Alsing2024}.}
    \label{err:Fig:z0MstarSFR}
\end{figure}

\begin{figure*}
    \centering
    \includegraphics[width=0.7\linewidth]{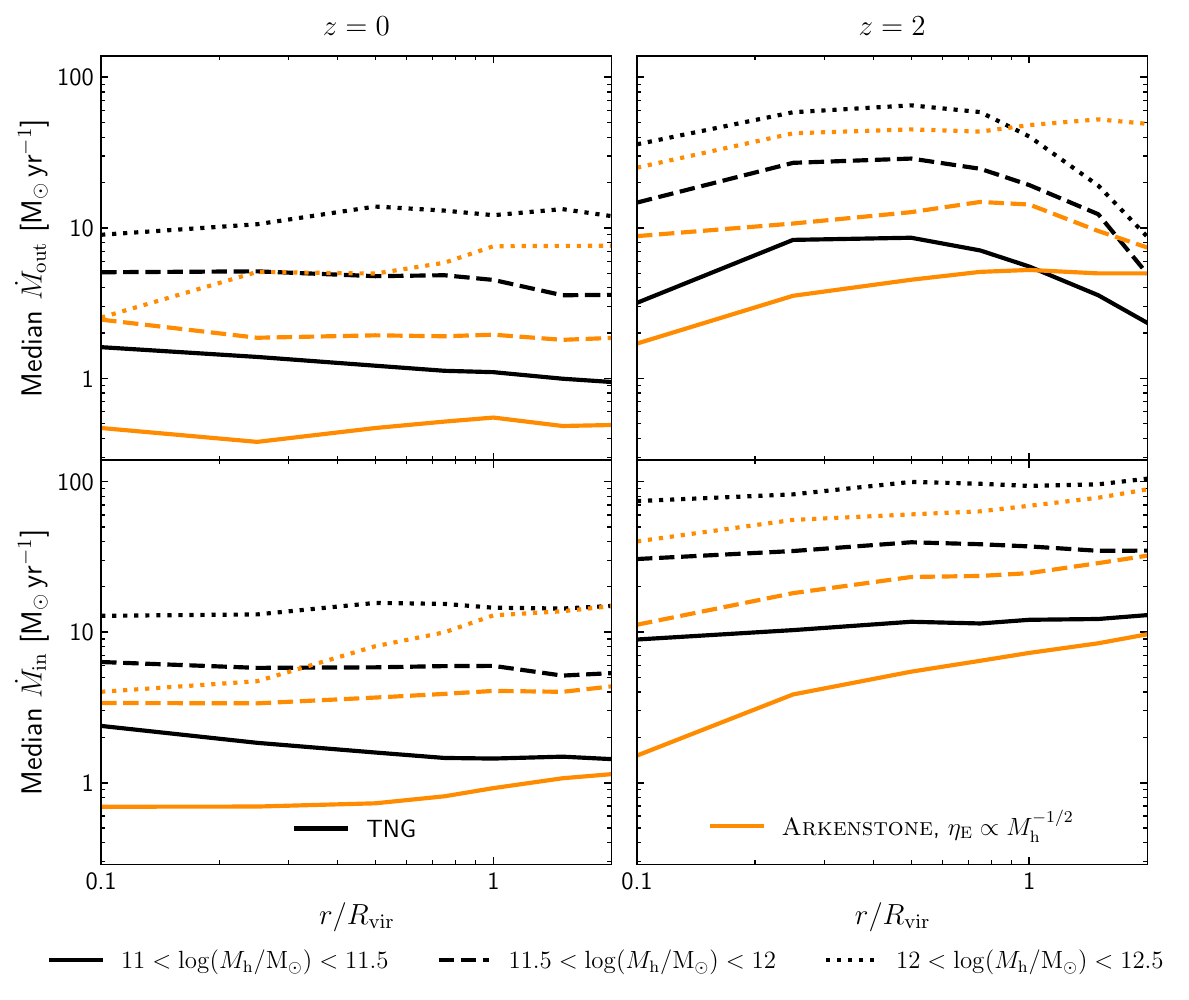}
    \caption{Corrected fig.~12 of \citet{err:Bennett2025}, showing the median mass outflow and inflow rates as a function of radius, binned in three different halo mass bins.}
    \label{err:Fig:inflowoutflow}
\end{figure*}

\begin{figure}
    \centering
    \includegraphics[width=0.8\linewidth]{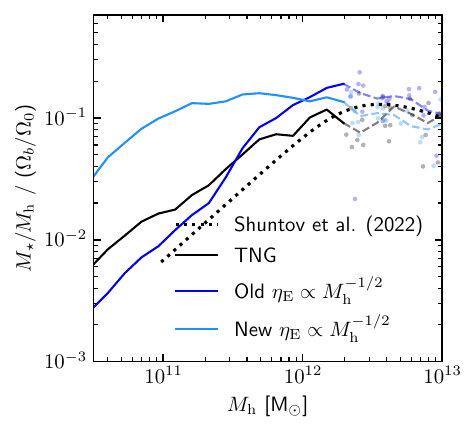}
    \caption{Corrected bottom panel of fig.~13 of \citet{err:Bennett2025}, showing the stellar mass to halo mass relation at $z=2$ normalised by the cosmic baryon fraction. Observational comparison data is from \citet{err:Shuntov2022}.}
    \label{err:Fig:z2SMHM}
\end{figure}

\begin{figure}
    \centering
    \includegraphics[width=0.8\linewidth]{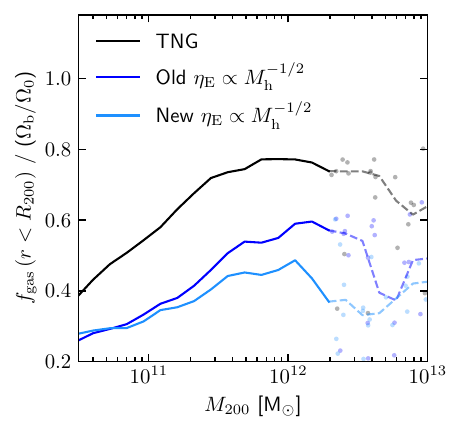}
    \caption{Corrected bottom panel of fig.~14 of \citet{err:Bennett2025}, showing the median gas fractions within $R_{200}$ at $z=2$ normalised by the cosmic baryon fraction.}
    \label{err:Fig:z2_mstar_fgas}
\end{figure}

\begin{figure}
    \centering
    \includegraphics[width=0.8\linewidth]{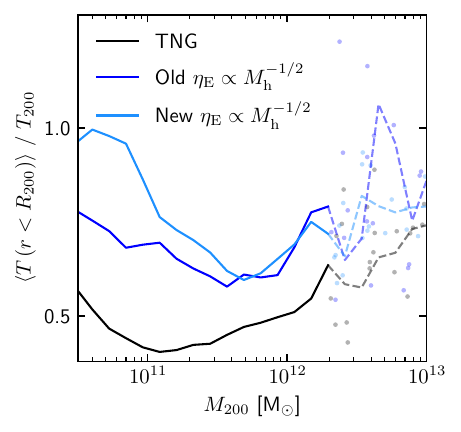}
    \caption{Corrected bottom panel of fig.~15 of \citet{err:Bennett2025}, showing the mass-weighted mean CGM temperatures within $R_{200}$ at $z=2$ normalised by the virial temperature.}
    \label{err:Fig:z2_mstar_TCGM}
\end{figure}

\begin{figure}
    \centering
    \includegraphics[width=0.8\linewidth]{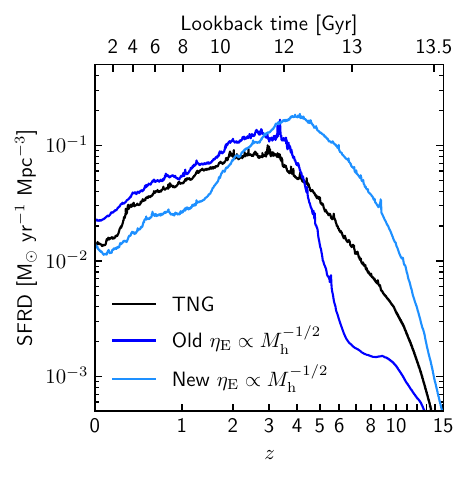}
    \caption{Corrected bottom panel of fig.~16 of \citet{err:Bennett2025}, showing the global SFR density in the simulation.}
    \label{err:Fig:SFRD}
\end{figure}

Fig.~\ref{err:Fig:DensMap} and \ref{err:Fig:DensRatioMap} show the gas surface density map of the new \textsc{Arkenstone} simulation box, and the ratio of this to the TNG box, respectively. These are very similar to the original versions.

In Fig.~\ref{err:Fig:z0SMF} we show the stellar mass function, and in Fig.~\ref{err:Fig:SMHM} we show the stellar mass to halo mass (SMHM) relation normalised by the cosmic baryon fraction, both at $z=0$, for TNG, the old, and the new \textsc{Arkenstone} simulations. 
We can see our new simulation in the light blue gives a reasonable match to the observational data \citet{err:Bernardi2017,err:Behroozi2019,err:Driver2022}, with a better match at intermediate masses compared to the old \textsc{Arkenstone} run, but slightly higher stellar masses at the low mass end. We note that, if we had adopted the original energy scaling (i.e. lower $\eta_E$ normalization) with the corrected code, the predicted stellar masses at $z=0$ (particularly in the dwarf regime) would have been much too high -- this simulation was started but found to be prohibitively expensive to run to $z=0$. 

The updated black hole to stellar mass relation is in Fig.~\ref{err:Fig:mstar_mbh} and is compared to the relation from \citet{err:KormendyHo2013}. We do find a slight difference in our corrected run compared to the old \textsc{Arkenstone} run, although, as discussed in the main paper, the black hole models have not been re-calibrated after introducing \textsc{Arkenstone}, and the interaction between the models will be studied in future work. 

In Fig.~\ref{err:Fig:Rgal} we show the relationship between stellar mass and stellar half mass radii, and again we see results comparable to observations \citep{err:Behroozi2022}. Galaxies are, on average, more compact in \textsc{Arkenstone} than TNG - the same conclusion we reached in the original paper.

The gas and baryon fractions at $z=0$, shown in Fig.~\ref{err:Fig:z0_mstar_fgas} and Fig.~\ref{err:Fig:z0_mstar_fbary} respectively, and the average CGM temperature in Fig.~\ref{err:Fig:z0_mstar_TCGM}, are largely similar in our new \textsc{Arkenstone} run, with only a slight reduction in gas and baryons in lower mass haloes. 

The corrected Figs.~\ref{err:Fig:z0MstarSFR} and \ref{err:Fig:inflowoutflow} have the same trends as the original paper, but with the effect of preventative feedback demonstrated more clearly -- strengthening our original conclusions. Fig.~\ref{err:Fig:z0MstarSFR} shows a clear correlation between star formation rate and CGM gas fraction, with galaxies lying below the main sequence having a depleted gaseous halo. The lower inflow (and outflow) rates compared to TNG in Fig.~\ref{err:Fig:inflowoutflow} also demonstrate preventative feedback in action, with lower inflowing mass fluxes at smaller radii for both $z=0$ and $z=2$.

The overestimation of thermal energy injection in our old runs at high redshift means the early behaviour of our new simulations is notably different, with significantly more stars formed at earlier times. We show how this impacts the $z=2$ SMHM relation in Fig.~\ref{err:Fig:z2SMHM}, where we can see a corresponding overproduction of stars in low mass haloes at this redshift, even though the stellar masses are reasonable by $z=0$. The gas fractions (Fig.~\ref{err:Fig:z2_mstar_fgas}) and average CGM temperatures (Fig.~\ref{err:Fig:z2_mstar_TCGM}) are comparable for the \textsc{Arkenstone} runs, and the relative difference to TNG is unchanged. 

The early enhancement of star formation in our new \textsc{Arkenstone} run is shown most clearly in the global SFR of our old and new simulations in Fig.~\ref{err:Fig:SFRD}. We will explore the impact of our changes at high redshift in detail in future work. However the strong implication from our new runs is that, for preventative feedback alone, dependence on dark matter velocity dispersion is insufficient to regulate high redshift star formation in low mass galaxies. There are several potential solutions to this -- early feedback could be more ejective, for example, or there could be a dependence of wind loadings on metallicity or ISM pressure that strengthens feedback at high redshift. These are scenarios we can investigate with the full \textsc{Arkenstone} model in future work. 

Notwithstanding the changes at high redshift, by $z=0$ we can reasonably recover our previous results with a global energy loading increase of around a factor of 2-3 and the removal of the cap at $\eta_\mathrm{E} = 1$. Importantly, the conclusions of the original paper remain unchanged, modulo the increase in energy loading described above. We note that despite this increase over the quoted \textsc{Arkenstone} values in the original paper, our new simulations still use a lower energy loading than TNG throughout the simulation. We further point out that while our new loadings exceed the nominal amount of energy expected to be released from supernovae at low halo masses, the calculation of this energy is uncertain (especially when considering the impact of pre-supernova feedback) and varies significantly between simulations.

\section*{Acknowledgements}
This work was supported by the Simons Collaboration on “Learning the Universe”. New computations were performed on Anvil, at Purdue University (accessed through allocation PHYS240043 from the Advanced Cyberinfrastructure Coordination Ecosystem: Services \& Support (ACCESS) program, which is supported by National Science Foundation grants \#2138259, \#2138286, \#2138307, \#2137603, and \#2138296); and Frontera (accessed through Leadership Resource Allocation AST20007), at the Texas Advanced Computing Center \citep{err:Frontera,err:Anvil,err:ACCESS}. 
GLB acknowledges support from the NSF (AST-2108470, AST-2307419), NASA TCAN award 80NSSC21K1053, and the Simons Foundation. The Flatiron Institute is supported by the Simons Foundation. 
This work made use of the \textsc{numpy} \citep{err:Numpy}, \textsc{scipy} \citep{err:SciPy}, and \textsc{matplotlib} \citep{err:Matplotlib} \textsc{python} packages.

\section*{Data Availability}
The data used in this work will be shared on reasonable request to the corresponding author.



\bsp
\label{lastpage}
\end{document}